\documentclass[11pt,a4paper]{article}
\pdfoutput=1
\usepackage[dvispsname]{xcolor}
\usepackage{jheppub}
\usepackage{epstopdf}
\usepackage{graphicx}
\usepackage{epsfig}
\usepackage{dcolumn}  
\usepackage{bm}    
\usepackage{amssymb} 
\usepackage{amsmath,bm}
\usepackage{amsfonts}    
\usepackage{slashed}  
 \usepackage{relsize}
\usepackage[mathscr]{euscript}
\usepackage{epsfig}
\usepackage{bbm}
\usepackage[utf8]{inputenc} 
\usepackage{euscript}
\usepackage{mathtools}
\usepackage[sf,isu]{caption}

 \preprint{IMSC/2016/FEB/8/1 \\ { \hspace*{12.3cm}TIFR/TH/16-04}}

\definecolor{labelkey}{rgb}{0.0,0.2706,0.4941}

\hypersetup{
	pdfstartview={FitH},    
	pdftitle={Higher-point conformal blocks and entanglement entropy in heavy states},    
	pdfauthor={Pinaki Banerjee, Shouvik Datta, Ritam Sinha},     
	colorlinks=true,       
	linkcolor=blue,          
	citecolor=blue!59!black! ,        
	filecolor=magenta,      
	urlcolor=blue           
}

\title{Higher-point conformal blocks and entanglement entropy in heavy states}

\author[s]{Pinaki Banerjee}
\author[t]{\!, Shouvik Datta}
\author[u]{\! and Ritam Sinha}

\affiliation[s]{\vspace{.3cm} Institute of Mathematical Sciences, \\
	CIT Campus, Taramani, \\ Chennai 600113, 	India. \\ \vspace{-.3cm}
	}
\affiliation[t]{Institut f\"{u}r Theoretische Physik,\\ ETH Z\"{u}rich, \\ 
	CH-8093 Z\"{u}rich, Switzerland.\\  \vspace{-.3cm}
	}
\affiliation[u]{Department of Theoretical Physics, \\
	Tata Institute of Fundamental Research,\\ Mumbai 400005, India.\\  \vspace{.1cm}
	}

\emailAdd{pinakib@imsc.res.in}
 \emailAdd{shouvik@itp.phys.ethz.ch}
  \emailAdd{ritam@theory.tifr.res.in}

\abstract{
	We consider conformal blocks of two heavy operators and an arbitrary  number of light operators in a (1+1)-$d$ CFT with large central charge. Using the monodromy method, these higher-point  conformal blocks are shown to factorize into products of 4-point conformal blocks in the heavy-light limit for a class of OPE channels. This result is reproduced by considering suitable worldline configurations in the bulk conical defect geometry.
	We apply the CFT results to calculate the entanglement entropy  of an arbitrary number of  disjoint intervals for heavy states. The corresponding holographic  entanglement entropy calculated
	 via the minimal area prescription precisely matches these results from CFT. Along the way, we briefly illustrate the relation of these conformal blocks to Riemann surfaces and their associated moduli space.
}

\begin{document}
 \maketitle

 \makeatletter
 \g@addto@macro\bfseries{\boldmath}
 \makeatother
 
 \def\th{\tilde{h}}
 \def\oh{\mathscr{O}_H}
 \def\ol{\mathscr{O}_L}
 \def\cX{\mathcal{X}}
 \def\beq{\begin{equation}}
 \def\eeq{\end{equation}}
\def\cB{\mathcal B}
\def\cZ{\mathcal N}
\def\cS{\mathcal S}
\def\bea{\begin{eqnarray}}
\def\eea{\end{eqnarray}}
\def\nn{\nonumber}
\def\pd{\partial}
\def\Re{R\'{e}nyi }
\def\l1{{\text{1-loop}}}
\def\uy{u_y}
\def\ur{u_R}
\def\o{\mathcal{O}}
\def\Cl{{{cl}}}
\def\bz{{\bar{z}}}
\def\by{{\bar{y}}}
\def\bX{\bar{X}}
\def\im{{\text{Im}}}
\def\re{{\text{Re}}}
\def\cn{{\text{cn}}}
\def\sn{{\text{sn}}}
\def\dn{{\text{dn}}}
\def\K{\mathbf{K}}
\def\n1{\Bigg|_{n=1}}
\def\fin{{\text{finite}}}
\def\R{{\mathscr{R}}}
\def\one{{(1)}}
\def\zero{{(0)}}
\def\n{{(n)}}
\def\tr{\text{Tr}}
\def\T{\mathcal{T}}
\def\TT{\tilde{\mathcal{T}}}
\def\O{\mathcal{O}}
\def\cN{\mathcal{N}}
\def\P{\Phi}
\def\W{{\tilde{W}}}
\def\T{{\tilde{T}}}
\def\I{\ \mathbf{\mathcal{I}}}
\def\ln{\boldsymbol{\colon}\hspace{-.13cm}}
\def\rn{\hspace{-.08cm}\boldsymbol{\colon}\hspace{-.05cm}}
\def\see{S_{\text{EE}}^{(2)}}
\def\rp{\rho_P}
\def\rq{\rho_Q}
\def\xpp{x^+_P}
\def\xpq{x^+_Q}
\def\xmp{x^-_P}
\def\xmq{x^-_Q}
\def\L{\, \mathbf{\mathsf{\Lambda}}}
\def\cc{\textsf{c.c.}}
\def\dl{\partial}
\def\la{\Bigg{\langle}}
\def\ra{\Bigg{\rangle}}
\def\bra#1{{\langle}#1|}
\def\obra#1{{\langle}\overline{#1}|}
\def\ket#1{|#1\rangle}
\def\oket#1{|\overline{#1}\rangle}
\def\bbra#1{{\langle\langle}#1|}
\def\kket#1{|#1\rangle\rangle}
\def\vev#1{\langle{#1}\rangle}
\def\cM{\mathcal{M}}
\def\cH{\mathcal{H}}
\def\red#1{\textcolor{red!70!black}{#1}}
\def\blue#1{\textcolor{blue}{#1}}
\def\bx{\bar{x}}
\def\bz{\bar{z}}
\def\hs{H_{\sigma}}
\def\hp{h_{\Phi}}
\def\tz{\tilde{z}}
\def\tbz{\bar{\tilde{z}}}
\def\cF{\mathcal{F}}
\def\ep{\epsilon}
\def\ein{{(1)}}
\def\zero{{(0)}}
\def\kay{{(k)}}
\def\zwei{{(2)}}
\def\drei{{(3)}}
\def\fear{{(4)}}
\def\funf{{(5)}}
\def\sechs{{(6)}}
\def\nbrac{{(N)}}
\def\cR{\mathcal{R}}

\def\f{\frac}
\def\tep{\tilde{\epsilon}}
\def\gap#1{\vspace{#1 ex}}
\def\be{\begin{equation}}
\def\ee{\end{equation}}
\def\bal{\begin{array}{l}}
\def\ba#1{\begin{array}{#1}}  
\def\ea{\end{array}}
\def\bea{\begin{eqnarray}}
\def\eea{\end{eqnarray}}
\def\beas{\begin{eqnarray*}}
\def\eeas{\end{eqnarray*}}
\def\del{\partial}
\def\eq#1{(\ref{#1})}
\def\fig#1{Fig.\,\ref{#1}} 
\def\re#1{{\bf #1}}
\def\bull{$\bullet$}
\def\nn{\\\nonumber}
\def\ub{\underbar}
\def\nl{\hfill\break}
\def\bibi{\bibitem}
\def\vev#1{\langle #1 \rangle} 
\def\mattwo#1#2#3#4{\left(\begin{array}{cc}#1&#2\\#3&#4\end{array}\right)} 
\def\tgen#1{T^{#1}}
\def\half{\frac12}
\def\floor#1{{\lfloor #1 \rfloor}}
\def\ceil#1{{\lceil #1 \rceil}}

\def\mysec#1{\gap1\ni{\bf #1}\gap1}
\def\mycap#1{\begin{quote}{\footnotesize #1}\end{quote}}

\def\Om{\mathbf{\Omega}}
\def\tom{\widetilde{\mathbf{\Omega}}}
\def\a{\alpha}
\def\b{\beta}
\def\l{\lambda}
\def\g{\gamma}
\def\sig{\sigma}
\def\eps{\epsilon}

\def\lan{\langle}
\def\ran{\rangle}
\def\righta{\to}

\def\nn{\nonumber}
\def\bit{\begin{item}}
\def\eit{\end{item}}
\def\benu{\begin{enumerate}}
\def\eenu{\end{enumerate}}

\def\tr{{\rm tr}}

\def\bx{\bar{x}}
\def\bz{\bar{z}}
\def\hs{H_{\sigma}}
\def\hp{h_{\Phi}}
\def\tz{\tilde{z}}
\def\tbz{\bar{\tilde{z}}}

\def\lb{\lbrace}
\def\rb{\rbrace}


\def\bT{\bar T}
\def\cO{{\mathcal O}}
\def\bl{{\bf L}}

\def\la{\langle}
\def\ra{\rangle}
\def\mapconfig{\Om_i \mapsto \lb (p,q)\rb}

\def\cA{\mathcal{A}}

\def\red#1{\textcolor{black}{#1}}
\section{Introduction}
Much of the power and appeal of holography \cite{Maldacena:1997re} rests on features of conformal field theories which find natural analogues in gravity. There are several of these features of holographic field theories which are universal and can be captured without reliance to a specific theory. Examples of these include thermodynamic features like the Cardy formula \cite{Cardy:1986ie} and entanglement entropy \cite{Calabrese:2004eu,Ryu:2006bv}. Moreover, it is of substantial interest to identify and explore the `universality class' of  CFTs which admit a holographic description \cite{Heemskerk:2009pn,ElShowk:2011ag,Hartman:2014oaa,Keller:2014xba,Haehl:2014yla,Belin:2014fna,Benjamin:2015hsa,Benjamin:2015vkc}.  

The evaluation of correlation functions by decomposing into conformal blocks is a minimalistic and powerful approach to extract  very general features of CFTs \cite{Ferrara:1971vh,zamo0,zamo1,Dolan:2003hv}. This direction,  recently harnessed by the conformal bootstrap programme,  has led to strong results on anomalous dimensions of operators \cite{ElShowk:2012ht,Rattazzi:2008pe,Rychkov:2015naa} and bounds on central charges \cite{Rattazzi:2010gj}. 
 In the AGT correspondence,  conformal blocks of Liouville theory (or more generally Toda theories)   are related to the instanton partition functions of 4-dimensional $\cN=2$ SCFTs \cite{Alday:2009aq,Wyllard:2009hg}.
 
Conformal blocks also play an important role in the context of holography since they serve as the CFT detectors of  bulk locality and  gravitational scattering \cite{Heemskerk:2009pn,Heemskerk:2010ty,Penedones:2010ue,ElShowk:2011ag,FKW,Jackson:2014nla,Maldacena:2015iua}.
Conjunctively, there has also been very strong evidence that conformal blocks are intimately related to geodesics in AdS \cite{Hijano:2015qja,Hijano:2015zsa,Hijano:2015rla,Alkalaev:2015wia,Alkalaev:2015lca,Alkalaev:2015fbw}. 
One of the important objects in this context, for a 2$d$ CFT, is the correlator of two heavy operators with two light operators \cite{FKW,Hijano:2015qja,Hijano:2015zsa,Hijano:2015rla,Alkalaev:2015wia,Alkalaev:2015lca,Alkalaev:2015fbw,Fitzpatrick:2015zha, Fitzpatrick:2015foa}\footnote{See references \cite{Litvinov:2013sxa,Perlmutter:2015iya,Chang:2015qfa,Beccaria:2015shq,Fitzpatrick:2015dlt,Headrick:2015gba,Menotti:2012wq,Menotti:2014kra,Menotti:2016jut} for further interesting aspects of conformal blocks in 2$d$ CFTs.}. As $c\to \infty$, the ratio of the conformal dimension with the central charge remains fixed for heavy operators, whilst that of light operators is much smaller than unity.  One can think of these heavy operators being responsible for creating an excited state after a global quench \cite{Caputa:2014vaa,Hartman2}. On the gravity side, this excited state corresponds to the conical defect background \cite{Hartman2}. It has been shown that the   conformal block of this correlator is precisely reproduced from holography from an appropriate worldline configuration in this bulk geometry \cite{Hijano:2015qja,Hijano:2015rla}. Moreover, the correlation functions in this  excited state mimic thermal behaviour if the conformal dimension of the heavy operator exciting the state is greater than $c/24$ \cite{Fitzpatrick:2015zha,Fitzpatrick:2015foa}. This is an  example of a pure microstate (with a sufficiently high energy eigenvalue) behaving effectively like a mixed state being a part of the thermal ensemble.

In this work, we evaluate  conformal blocks of two heavy operators and arbitrary  number $(m)$ of light operators. We work in the heavy-light approximation and utilize the monodromy method to derive the ($m+2$)-point conformal block. We expand the correlation function in a basis which involves pairwise fusion of the light operators (see \fig{intro-fig}). In the strict heavy-light limit and at large central charge, we show that,  for this class of OPE channels :
\begin{itemize}
	\item The conformal block having an even number of light operators and two heavy operators factorizes into a product of 4-point conformal blocks of two heavy and two light operators. 
		\item The conformal block having an odd number of light operators and two heavy operators factorizes into a product of 4-point conformal blocks of two heavy and two light operators and a 3-point function involving one light and two heavy operators.
\end{itemize}
Our monodromy analysis is developed mostly based on the work \cite{Alkalaev:2015lca} which made use of  the accessory parameters of the 4-point conformal block as a seed solution in order to obtain those for the 5-point conformal block.    Although the factorization we observe  is special to the OPE channel configurations we have chosen,  the conformal blocks in other bases can be related to ours by performing linear operations. Furthermore, since the correlator itself is a basis independent object, all bases of conformal blocks are on an equal footing. We shall demonstrate this picture by using the correspondence relating CFT correlators to punctured Riemann surfaces. 
\begin{figure}[!b]
	\centering
	\begin{tabular}{c}
		\includegraphics[width=.88\textwidth]{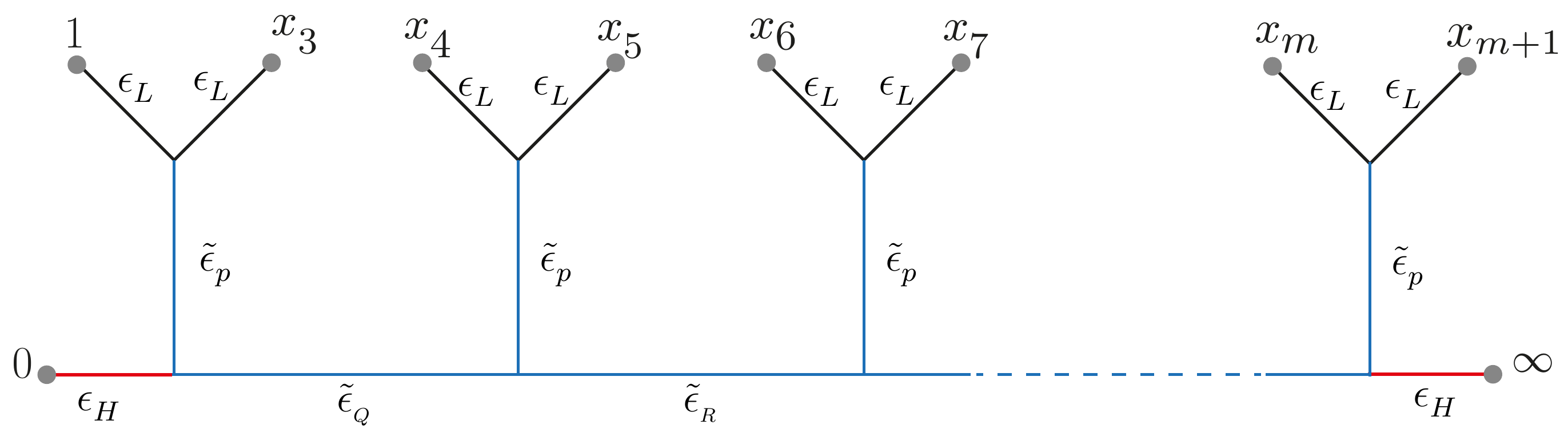} 
	\end{tabular}
	\caption{\small The OPE channel which we shall consider  here for the conformal block of an even number light operators and two heavy operators. The conformal dimensions are scaled as $\ep_i =6h_i/c$. Note the pairwise fusion of the light operators into  operators of conformal dimension $\th_p$ which are same in the intermediate channels shown in the figure -- this provides a major simplification for the monodromy analysis.}
	\label{intro-fig}
\end{figure}

Our results for  conformal blocks from the CFT are reproduced from the bulk by considering suitable generalizations of the worldline configurations considered in \cite{Hijano:2015rla}. The choice of OPE channels in the CFT are in one-to-one correspondence to geodesic configurations in the bulk. This implies that these higher-point conformal blocks can be fully recast in terms of  bulk quantities. This outcome nicely fits within the notion of emergence of locality from a conformal field theory \cite{ElShowk:2011ag} and serves as an explicit demonstration of the same not only for  higher-point correlation functions but also for non-vacuum states. 
It was also shown previously that correlation functions of free theories can be rewritten in terms of closed string amplitudes in AdS  \cite{Gopakumar:2003ns,Gopakumar:2004qb,Gopakumar:2005fx}. 
Although our correlator is in a very different regime of the parameter space of couplings, it bears in the same spirit the pertinent analogue of admitting an Einstein gravity description instead of the stringy one for free theories. 

This circle of ideas finds a natural home in the context of entanglement entropy of heavy states \cite{Hartman2}. The light operators then correspond to twist operators, with conformal dimension $c/24\, (n-1/n)$, (which implement the replica trick) in the limit $n \to 1$, where $n$ is the replica index \cite{Calabrese:2004eu}. One can then utilize the higher-point conformal block to obtain the entanglement entropy of an arbitrary number of disjoint intervals. Our choice of monodromy contours, similar to those used in \cite{Hartman1,Faulkner}, are well-suited to computing entanglement entropy. 
Furthermore, these results can be straightforwardly used to evaluate the mutual information of two  disjoint intervals, $\cA$ and $\cB$, which is defined as 
\begin{equation*}
I(\cA,\cB) =  S(\cA) + S(\cB) - S\left(\cA\cup \cB\right) .
\end{equation*}
On the holographic side, the Ryu-Takayanagi prescription \cite{Ryu:2006bv} instructs us to calculate minimal areas (or geodesics for the case of 3$d$ gravity). Our CFT results for entanglement entropy of disjoint intervals agree precisely with that obtained from the bulk using the minimal area prescription. Furthermore, we show that for two or more disjoint intervals, there are multiple geodesic configurations which are possible in the bulk and these are in one-to-one correspondence with various OPE channels or monodromy contours which one chooses to consider in the CFT. 


The outline of this paper is as follows. In Section \ref{sec:2}, we introduce the correlation function whose conformal block we wish to evaluate and specify the regime of validity of our analysis. The relation  of CFT correlators with punctured Riemann surfaces is reviewed in this section.  Section \ref{sec:3} contains the explicit evaluation of the conformal block. We consider the 5- and 6-point conformal blocks before generalizing the result to arbitrary odd- and even-point blocks. Specializing to the case of light primaries being twist operators, we evaluate the entanglement entropy and mutual information in Section \ref{sec:4}. In Section \ref{sec:5}, we consider worldline configurations in the bulk and reproduce the CFT result for odd- and even-point blocks. This section also contains the analysis of the holographic entanglement entropy using the Ryu-Takayanagi formula.  
We describe the moduli space of the conformal block in Section \ref{sec:7}. Finally, Section \ref{sec:8} has our conclusions along with some future directions.



\def\so{\mathscr{O}}
\section{On heavy-light correlators and Riemann surfaces}\label{sec:2}
Consider the following $p$-point correlation function of primary operators  $\so_i$, each located at the points $(z_i,\bz_i)$ on a plane. By inserting $(p-3)$ number of   complete sets of states equivalent to the identity (\fig{intro-fig}), we can expand in terms of the conformal partial waves as 
\begin{align}
&\la \so_1(z_1,\bz_1) \so_2(z_2,\bz_2)  \cdots \so_p(z_p,\bz_p)    \ra  
= \sum_{\lbrace\tilde{h}_i \rbrace}d_{\lb \tilde{h}_i\rb} \cF (z_i,h_i,\tilde{h}_i) \bar\cF (\bz_i,h_i,\tilde{h}_i).
\end{align}
The complete set of states $\ket{\tilde{h}_i}$, running in the  $(p-3)$  intermediate channels,  is labelled in terms of representations of the Virasoro algebra which includes the primaries as well as  their descendants. 
Here, $d_{\lb \tilde{h}_i\rb}$ are constructed out of the structure constants $c_{ijk}$ of the algebra of operators in the theory. In what follows, we shall be interested in the following $(m+2)$-point correlator of two heavy operators and $m$ light operators\footnote{In this section and the next we move back and forth between using $p$ and $m$. $p=m+2$.}
\begin{align}\label{our-corr}
\langle \oh(z_1,\bz_1) \  \prod_{i=2}^{m+1}\ol (z_i,\bz_i) \  \oh (z_{m+2},\bz_{m+2})  \rangle  .
\end{align}

We shall work with CFTs, which in the $c\to \infty$ regime, admit a holographic description in terms of Einstein gravity.  We  make use of the property that, in this regime, the $p$-point conformal blocks are expected to exponentiate as \cite{zamo0,zamo1,zamozamo}
\begin{align}\label{block-exponentiation}
\cF_{(p)}(z_i,h_i, \tilde h_i) = \exp \left[-\frac{c}{6} f_{(p)}(z_i,\epsilon_i, \tilde\epsilon_i) \right].
\end{align}
Furthermore, the points $z_1$, $z_2$ and $z_{m+2}$, in \eqref{our-corr}, can be sent to $\infty$, $1$ and $0$ respectively via the projective transformation 
\begin{align}\label{pro-trans}
x_{i} = \frac{(z_{m+2} - z_i)(z_2-z_1)}{(z_{m+2}-z_2)(z_i-z_1)}.
\end{align}
Upto factors of the Jacobians arising from usual rules of conformal transformations of primary operators, the correlator of interest is now expressed in terms of the cross-ratios, $x_i$, as
\begin{align}\label{our-corr-2}
\Big\langle \oh(\infty) \left[ \ol(1) \prod_{i=3}^{m +1}\ol (x_i) \right] \oh(0) \Big\rangle  .
\end{align}
We shall be interested in contributions of  Virasoro conformal blocks, $\cF_{(p)}(x_i,h_i, \tilde h_i)$, to the above correlation function. 
In addition to the $c \to \infty$ limit, we shall also work in the heavy-light limit, for which the dimensions of the operators scale as
\begin{align}\label{conf-dims}
  \ep_H = \frac{6h_H }{c} \sim \mathcal{O}(1) \ , \quad  \ep_L=\frac{ 6h_L }{ c }\ll 1  \ .
\end{align}
In words, the ratio of the conformal dimension with central charge of the heavy operators remains fixed and is of order-one in the $c\to \infty$ regime whilst that of the light operators much lesser than unity\footnote{In Section \ref{sec:4}, we specialize to the case of the light operators being twist operators with conformal dimension $c/24\, (n-1/n)$ in the limit $n \to 1$. Here, $n$ is the index for the number of replicas. }.  By referring to this as the  `heavy-light limit',  we mean to consider contributions  to the conformal block to the leading order in $\ep_L$.

The four-point version of the above conformal block was considered in \cite{FKW} and the five-point in \cite{Alkalaev:2015lca}. The conformal block was used to calculate the entanglement entropy of a single interval in heavy states in \cite{Hartman2}.

It is important to note, that there are several choices of OPE channels along which one can expand a CFT correlation function. Each of these channels correspond to different basis choices of the conformal blocks to rewrite the correlator. The correlation function is   a single-valued real analytic function of the coordinates $z_i$ and $\bz_i$. However, this is not true for the conformal blocks themselves (due to presence of branch cuts) and the correlation function is, therefore, independent of the basis of conformal blocks. It can be shown that conformal blocks in different bases are related by the linear operations of  braiding, fusion and modular transformation  \cite{Moore:1988qv,Moore:Poly,Moore:Lec,Jackson:2014nla}. These operations have finite-dimensional representations for rational CFTs but  can, nevertheless, be performed on conformal blocks of a generic CFT since these are duality transformations purely arising from associativity of OPEs (crossing-symmetry) and modular invariance\footnote{See, for example, \cite{Jackson:2014nla} which   explicitly constructs the duality transformations in terms of quantum $6j$-symbols for Liouville theory. Quite intriguingly, the authors also show that the braiding matrix  is related to gravitational scattering amplitudes.}. In this paper, we shall work in a specific basis in which light operators fuse in pairs. As we shall show, this basis admits a generalization to higher point conformal blocks and is geared towards the analysis of  entanglement entropy of disjoint intervals \cite{Hartman1,Faulkner}. The OPE channel considered for the 5-point function in \cite{Alkalaev:2015lca} is different from the one we are about to use but is, however, related  to ours by a series of fusion operations\footnote{See Fig.~18 of \cite{Moore:1988qv} to relate the basis used in \cite{Alkalaev:2015lca} to the one used here.}. 

 The statements above can be manifestly portrayed if one associates, a  Riemann sphere with $p$ punctures, $\Sigma_{0,p}$,  with a $p$-point CFT correlation function on the plane  \cite{Moore:1988qv,Moore:Poly,Moore:Lec,Zwiebach:1990nh,Jackson:2014nla}. Strictly speaking, there is a vector space of conformal blocks $\cH(\Sigma)$ associated to every Riemann surface $\Sigma$ \cite{Friedan:1986ua}. The different bases of conformal blocks (or OPE channel choices) are the various ways in which this Riemann surface can be sewn from 3-holed spheres (or as is graphically called `a pair of pants'). Stated differently, the decomposition of the correlator into conformal blocks is equivalent to the pant-decomposition of the punctured sphere. The intermediate channels in the conformal blocks correspond to the states passing through the sewed holes. The number of intermediate sewings necessary  for the $p$-punctured sphere is $(p-3)$.


  Moreover, a $p$-point CFT correlation function on the plane  is 
  related to a moduli space,  $\cM_{0,p}$, (corresponding to the Riemann surface $\Sigma_{0,p}$)  \cite{Friedan:1986ua,Moore:1988qv,Zwiebach:1990nh}. For our correlator \eqref{our-corr-2} on the sphere, this is $\cM_{0,m+2}$. There are $(m-1)$ complex moduli formed by the cross-ratios $\lbrace x_i \rbrace$. This picture will turn out to be relevant later on in Section \ref{sec:7}.


\def\hpsi{\hat\psi}
\section{Monodromy problem for  conformal blocks}\label{sec:3}
 
 In this section we briefly review the monodromy method to evaluate the Virasoro conformal block. The discussion closely follows \cite{FKW,deBoer:2014sna}. We then proceed to use it for calculating  5- and 6-point conformal blocks and then generalize to blocks with an arbitrary number of light operator insertions. 
 
 Let us consider the correlation function we started with, $\la \so_1(z_1) \so_2(z_2) \so_3(z_3) \cdots \so_p(z_p) \ra $. As mentioned before, in order to decompose this $p$-point correlator into a sum of products of 3 point functions we need to insert $p-3$ resolutions of identity. In terms of these intermediate states, a typical conformal block would read as\footnote{Henceforth, we focus attention on the holomorphic part.}
 \begin{align}\label{break-3pt}
\cF_{(p)}(z_i,h_i,\tilde{h}_i)=\la  \so_1(z_1) \so_2(z_2)\ket{\alpha} \, \bra{\alpha}\so_3(z_3)  \ket{\beta}\, \bra{\beta} \so_4(z_4) \ket{\gamma}\cdots \bra{\zeta} \so_{p-1}(z_{p-1})\so_p(z_p)  \ra  .
 \end{align}
 Now, one can insert  into this conformal block an additional operator, $\hpsi(z)$, whose conformal dimension remains fixed in the $c \to \infty$ limit. This defines the following quantity 
 \begin{align}
 \Psi(z,z_i) \equiv \la  \so_1(z_1) \so_2(z_2)\ket{\alpha} \, \bra{\alpha}\hpsi(z)\so_3(z_3)  \ket{\beta}\, \cdots \bra{\zeta} \so_{p-1}(z_{p-1})\so_p(z_p)  \ra  \nn .
 \end{align}
 It can then be argued \cite{FKW} that the insertion of $\hpsi$ changes the leading semi-classical behaviour of the conformal block just by multiplication of a wavefunction 
 \begin{align}\label{wavefn}
\Psi(z,z_i) = \psi(z,z_i) \cF_{(p)}(z_i,h_i,\tilde{h}_i) .
 \end{align}
 We can now choose that the operator $\hpsi$ obeys the null-state condition at level 2. 
  \be
\left[  L_{-2}-\f{3}{2(2 h_\psi +1)}L_{-1}^2\right]\ket{\psi}=0, \qquad \text{with,  }h_\psi \stackrel{c\to \infty}{=}-\f12-\f9{2c} .
  \ee
Acting $\Psi(z,z_i)$ by  $(L_{-2}-\f{3}{2(2 h_\psi +1)}L_{-1}^2)$, therefore, leads to
  \be
  \left[L_{-2}+\f{c}6 L_{-1}^2\right]\Psi (z,z_i)=0.
  \ee
  Translating this into a differential equation by using the differential operator realization of Virasoro generators, one arrives at a Fuchsian equation 
\begin{align}\label{monodromy-01}
\frac{d^2 \psi(z)}{dz^2} + T(z) \psi(z) =0, \ \  \quad \text{with,  \  } T(z)= \sum_{i=1}^{p} \left[ \frac{\ep_i}{(z-z_i)^2} + \frac{c_i}{z- z_i }   \right].
\end{align}
Here, $\ep_i=6h_i/c$ and $c_i$ are the accessory parameters related to the conformal block as 
\begin{align} \label{acc-para}
c_i = -  \frac{\pd f_{(p)}(z_i,\ep_i,\tep_i)}{\pd z_i}.
\end{align}The function $f_{(p)}$ is the same as the one appearing in the exponential in equation \eqref{block-exponentiation} and forms the essential ingredient of the conformal block. 
Equation \eqref{acc-para} implies that the following {integrability condition} should be satisfied
\begin{align}\label{integrability-01}
\frac{\pd c_i}{\pd z_j}  =   \frac{\pd c_j}{\pd z_i}.
\end{align}
The asymptotic behaviour $T(z) \sim 1/z^4$, at $z\to \infty$, imposes the conditions 
\begin{align}
\sum_{i=1}^{p} c_i =0 , \quad \sum_{i=1}^{p} (c_i z_i + \ep_i) =0 , \quad \sum_{i=1}^{p} (c_i z_i ^2+ 2\ep_iz_i) =0 .
\end{align}
We shall now work in the coordinate system $x_i$ as defined in \eqref{pro-trans} and also consider two heavy operators at $0$ and $\infty$ and $(p-2)$ light operators at $1,x_3,\cdots,x_{p-1}$. 
Using the above three relations, we can re-express $c_{1,2,p}$ in terms of other accessory parameters $c_{3,\cdots,p-1}$ and cross-ratios as  
\begin{align}\label{ci}
c_1 = \sum_{i=3}^{p-1} (x_i-1)c_i + (p-2)\, \ep_L, \qquad 
c_2 = -\sum_{i=3}^{p-1}x_i c_i - (p-2) \,  \ep_L , \qquad 
c_{p} &= 0 .
\end{align}
Substituting \eqref{conf-dims} and \eqref{ci} in the expression for $T(z)$ we get
\begin{align}\label{Tz-full}
T(z) = &\frac{\ep_H}{z^2} + \frac{\ep_L}{(z-1)^2} + \sum_{i=3}^{p-1} \frac{\ep_L}{(z-x_i)^2}  - \frac{(p-2)\,  \ep_L}{z(z-1)}  +\sum_{i=3}^{p-1} \frac{x_i(1-x_i)}{z(z-1)(z-x_i)} c_i   .
\end{align}
One can solve for the accessory parameters, $c_i$, by using the monodromy properties of the solution $\psi(z)$ around the singularities of $T(z)$. 
The main ingredient of the conformal block $f_{(p)}(x_i,\ep_i,\tep_i)$ can, in turn, be obtained from the accessory parameters upon integrating, $c_i = - \pd f_{(p)}/\pd x_i$. Thinking of the  conformal block $\cF_{(p)}(x_i,h_i, \tilde h_i)$ as a partition function dominated by the saddle-point action $(c/6) f_{(p)}(x_i,\ep_i,\tep_i)$, the accessory parameters $c_i$ serve as  conjugate variables for the cross-ratios $x_i$\,\footnote{It was pointed out in \cite{Jackson:2014nla} that the corresponding Teichm\"{u}ller space inherits the complex structure $(x_i,\gamma_i)$ from the punctured Riemann sphere. The authors had also studied quantization arising from the conjugate variables $x_i$ and $\gamma_i$. The wavefunction $\Psi(z,z_i)$ in \eqref{wavefn} can then be thought to be living in the associated Hilbert space.}. As emphasized in \cite{Hijano:2015rla}, the saddle-point action dominating  the conformal block is closely related to the worldline action in AdS. The accessory parameter can then be identified as the momentum along the geodesic. 

\red{As mentioned earlier, we shall focus in the OPE channels in which the light operators fuse in pairs. Since we are interested in universal properties of CFTs which have a holographic dual as classical Einstein gravity, we shall assume that fusion channel of two light operators (the vertical channels in \fig{intro-fig}) is dominated by the exchange of the Virasoro identity block. This consists of operators  dual to graviton fluctuations in the bulk which have conformal dimensions much lesser than the light operators (whose conformal dimension scales as the central charge). An explicit example of this is that of twist operators, which become light in the $n \to 1$ limit and the contribution in their fusion channel is dominated by the identity block \cite{Asplund:2013zba}. The identity block includes the stress tensor and its descendants ($T, \, \pd T, \, TT, \pd^2 T, \cdots$) which have conformal dimensions which scale as $\mathcal{O}(1)$ as opposed to $\mathcal{O}(c)$ for the heavy and light operators. This is the only assumption on the spectrum we shall make in our following analysis. }

We shall now consider two warmup examples of the $5$-  and   $6$-point conformal blocks to gain some intuition before analyzing the arbitrary $p$-point case. Let us reiterate the notation for the number of insertions of operators
\begin{align*}
 \text{total number of operator insertions} &=p , \\
 \text{number of light operator insertions} &= m ,\\
 \text{number of heavy operator insertions} &= 2 .
\end{align*}
Therefore, $p=m+2$. The variable $n$ is reserved for the replica index which will appear in the context of entanglement entropy. 
Along the way, we also mention the procedure to be used and fix the notation and conventions further. 
\subsection{Warmup examples}
\label{subsec:warmup}
\subsection*{\textit{Warmup example I : 5-point conformal block}}
Let us consider the conformal block of the 5-point function
\begin{align}
\la \oh(\infty) \ol(1) \ol(x_3) \ol(x_4)  \oh(0) \ra .
\end{align}
This is the case of $m=3$ or $p=5$. 
\subsubsection*{Perturbative expansion of the monodromy equation}

We shall solve for the unknown accessory parameters $c_{3,4}$ in perturbation theory in the parameter $\ep_L$ which is the scaled conformal dimension of the light operator  \eqref{conf-dims}. 
We shall be following the procedure used in \cite{FKW,Alkalaev:2015lca,Hijano:2015rla}. However, as mentioned earlier, unlike \cite{Alkalaev:2015lca} which studied the 5-point block in much greater detail, our choice of monodromy contours will be different and we shall be working only up to linear order in $\ep_L$ which is the light parameter of all our light operators. 

The quantities in the monodromy equation \eqref{monodromy-01}  can be expanded implicitly  in powers of the light parameter, $\ep_L$, as 
\begin{align}\label{pert-sol}
\psi(z) &= \psi^{(0)}(z)   + \psi^{(1)}(z)   + \psi^{(2)}(z)  + \cdots, \nn \\
T(z) &= T^{(0)}(z)   + T^{(1)}(z)   + T^{(2)}(z)  + \cdots, \\
c_i (z) &= c_i^{(0)}(z)   + c_i^{(1)}(z)   + c_i^{(2)}(z)  + \cdots,  \quad \text{for }i=3,4,\dots,m+1. \nn 
\end{align}
For the case at hand, $m=3$. Following \cite{Alkalaev:2015lca}, we also assume that the expansion of the  accessory parameters starts at linear order in $\ep_L$ and hence $c_i^{(0)}=0$.  The equation \eqref{monodromy-01} at the first two orders is 
\begin{align}
&\frac{d^2 \psi^{(0)}(z)}{dz^2} + T^{(0)}(z) \psi^{(0)} (z) = 0, \label{mono-0}\\
&\frac{d^2 \psi^{(1)}(z)}{dz^2} + T^{(0)}(z) \psi^{(1)} (z) =  - T^{(1)}\psi^{(0)}(z) .\label{mono-1}
\end{align}
From \eqref{Tz-full}, the stress-tensor at these orders is
\begin{align}
T^{(0)} (z) = &\frac{\ep_H}{z^2}, \\
T^{(1)} (z) = & \frac{\ep_L}{(z-1)^2} + \frac{\ep_L}{(z-x_3)^2} +  \frac{\ep_L}{(z-x_4)^2} - \frac{3 \ep_L}{z(z-1)}   \\
& \quad 
+ \frac{x_3(1-x_3)}{z(z-1)(z-x_3)} c^{(1)}_3+ \frac{x_4(1-x_4)}{z(z-1)(z-x_4)} c^{(1)}_4 .\nn 
\end{align}
As in \cite{Alkalaev:2015lca}, we shall also supress the superscript of $c_i^{(1)}$ and simply call it $c_i$ since  the accessory parameters at the linear order are sufficient information for the heavy-light limit. In other words, we shall confine our attention to the linear order in $\ep_L$ which is the extreme heavy-light limit. 

The solution to the zeroth order ODE in \eqref{mono-0} is straightforward
\begin{align}\label{zeroth-order-sol}
\psi_\pm^{(0)} (z) = z^{ (1\pm \alpha)/2} , \quad \quad \alpha = \sqrt{1-4 \ep_H}.
\end{align}
Let us calculate the monodromy of this first order solution about the points $0$ and $\infty$. It can be easily seen that taking $z$ to $e^{2\pi i}z$, results in 
\begin{align}\label{monodromyat0}
\begin{pmatrix}
\psi_+^\zero (e^{2\pi i }z) \\
\psi_-^\zero (e^{2\pi i }z)
\end{pmatrix}= -\begin{pmatrix}
e^{\pi i \alpha} &0 \\
0 &e^{-\pi i \alpha} \\
\end{pmatrix}\begin{pmatrix}
\psi_+^\zero ( z) \\
\psi_-^\zero ( z)
\end{pmatrix}.
\end{align}
The $2\times2$ matrix above  is therefore the monodromy matrix for a contour containing the point $z=0$. In a similar fashion the monodromy around $z=\infty$ can also be seen by performing the transformation $y=1/z$
\begin{align}\label{monodromyatinf}
\begin{pmatrix}
\psi_+^\zero (e^{2\pi i }y) \\
\psi_-^\zero (e^{2\pi i }y)
\end{pmatrix}= -\begin{pmatrix}
e^{-\pi i \alpha} &0 \\
0 &e^{\pi i \alpha} \\
\end{pmatrix}\begin{pmatrix}
\psi_+^\zero ( y) \\
\psi_-^\zero ( y)
\end{pmatrix}.
\end{align}
Since $\alpha= \sqrt{1-4 \ep_H}$, the monodromies above detect the conformal dimensions of the heavy operators inserted at $0$ and $\infty$.

The first order corrections \eqref{mono-1} can be obtained by the standard method of variation of parameters. The Wronskian is $W(z)=\alpha$. We have
\begin{align}\label{mvp}
\psi^\ein_+ (z) &= \frac{1}{\alpha} \psi^\zero_+ (z) \int dz \psi^\zero_- (z) T^\ein(z) \psi^\zero_+ (z) - \frac{1}{\alpha} \psi^\zero_- (z) \int dz \psi^\zero_+ (z) T^\ein(z) \psi^\zero_+ (z), \nn \\
\psi^\ein_- (z) &= \frac{1}{\alpha} \psi^\zero_+ (z) \int dz \psi^\zero_- (z) T^\ein(z) \psi^\zero_- (z) - \frac{1}{\alpha} \psi^\zero_- (z) \int dz \psi^\zero_- (z) T^\ein(z) \psi^\zero_+ (z) .
\end{align}

\subsubsection*{Monodromy conditions}


$T^\ein(z)$ has three singular points at $1, \, x_3$ and $x_5$ { i.e.}~at the location of the light operators. Our choice of contours shall involve one contour enclosing a pair of points and another one enclosing the remaining single point. There are three such possibilities 
\begin{itemize}
   \item $\Om _1$ : $\gamma_1$ enclosing $\lb 1,x_3 \rb$ and $\gamma_2$ enclosing $\lb x_4 \rb$
    \item  $\Om _2$ : $\gamma_1$ enclosing $\lb 1,x_4 \rb$ and $\gamma_2$ enclosing $\lb x_3 \rb$
     \item  $\Om _3$ : $\gamma_1$ enclosing  $\lb x_3,x_4 \rb$ and $\gamma_2$ enclosing $\lb 1 \rb$
\end{itemize}
  These contour configurations are in one-to-one correspondence with the OPE channel along which one chooses to expand. For instance, for $\Om_1$, we consider the OPE,  $\ol(1)\ol(x_3)$. We shall elaborate on this further below.

The monodromy matrix upto first order in $\epsilon_L$ is
\begin{align}\label{mono-matrix-01}
\mathbb{M}({\gamma_k}) =  \mathbb{I}+ \begin{pmatrix}
I^\kay_{++}  & I^\kay_{+-} \\
I^\kay_{-+}  & I^\kay_{--}
\end{pmatrix},
\end{align}
where, the $I^\kay_{pq}$ are contour integrals
\begin{align}\label{mono-elements}
I^\kay_{++} = \frac{1}{\alpha} \, \oint_{\gamma_k} dz \,\psi^\zero_- (z) T^\ein(z) \psi^\zero_+(z) , \quad  I^\kay_{+-} = -\frac{1}{\alpha} \, \oint_{\gamma_k} dz \, \psi^\zero_+ (z) T^\ein(z) \psi^\zero_+ (z) \nn \\ 
I^\kay_{-+} = \frac{1}{\alpha} \, \oint_{\gamma_k} dz \,\psi^\zero_- (z) T^\ein(z) \psi^\zero_-(z) , \quad  I^\kay_{--} = -\frac{1}{\alpha} \, \oint_{\gamma_k} dz \, \psi^\zero_+ (z) T^\ein(z) \psi^\zero_- (z)
\end{align}
  Note that, $I^\kay_{++} = - I^\kay_{--} $.

The monodromy conditions we shall impose are \cite{FKW} 
\begin{align}\label{mono-def}
\widetilde{\mathbb{M}}(\gamma_k) = - \begin{pmatrix}
e^{+\pi i \Lambda}  & 0 \\
0 & e^{-\pi i \Lambda}  
\end{pmatrix}   \ , \qquad \Lambda = \sqrt{1-4 \tilde\ep_p} \ .
\end{align}
In words, the above equation means that the monodromy matrix picks up the conformal dimension, $\tilde{h}_p=c\,\tep_p/6$, of the operator $\so_p$ which arises upon fusing the operators living inside the contour. 
The tilde on $\mathbb{M}$ above denotes that this diagonal form of the  monodromy matrix is related by similarity transformations to \eqref{mono-matrix-01}. 
Comparing the eigenvalues of \eqref{mono-matrix-01} with that of $\widetilde{\mathbb{M}}(\gamma_k)$ we get the condition
\begin{align}\label{eigen-condition}
\mathcal{X}[\gamma_k]\equiv \left(I_{++}^\kay \right)^2 + I^\kay_{+-} I^\kay_{-+} &= -4 \pi^2 \tilde\ep_p^2   .
\end{align}
Here, we have defined $\cX [\gamma_k]$  as the monodromy condition for the contour $\gamma_k$ at the linear order in $\ep_L$. 
We shall now impose this condition for each of the contours in the configurations, $\Om_i$. It is worthwhile remarking at this point that, these configurations of monodromy contours are in one-to-one correspondence with the OPE channels. The residues provide information about the singular structure due to the operators residing within the contours. 

\subsubsection*{$\Om_1$ channel}
The $\Om_1$ channel --  \fig{3-1} -- corresponds to case when we consider the fusion of the light primaries $\ol(1)$ and $\ol(x_3)$. 
The monodromy conditions for this configuration is 
\begin{align}\label{51}
\cX[\gamma_1]=- \frac{4\pi^2  x_3^{-\alpha/2}}{\alpha}&[\ep_L(\alpha-1+(\alpha+1)x_3^\alpha)+(x_3^\alpha-1) x_3 c_3]    \nn \\ &\times [\ep_L(\alpha-2+(\alpha+2)x_3^\alpha)+(x_3^\alpha-1)(c_3x_3+c_4x_4)] \  = \ - 4\pi^2 \tilde\ep_p^2   , 
\end{align}
and 
\begin{align}
\cX[\gamma_2]=-4 \pi^2 \ep_L^2  = -4 \pi^2 \tilde\ep_q^2 .
\end{align}
The last equation above is fairly obvious. The residue merely picks up the conformal dimension of the single operator $\ol(x_4)$ living inside the contour $\gamma_2$. Hence, $\tep_q=\ep_L$, which is consistent with the conformal block diagram shown in \fig{3-1}.

\begin{figure}[!t]
	\centering
	\begin{tabular}{c}
		\includegraphics[width=.75\textwidth]{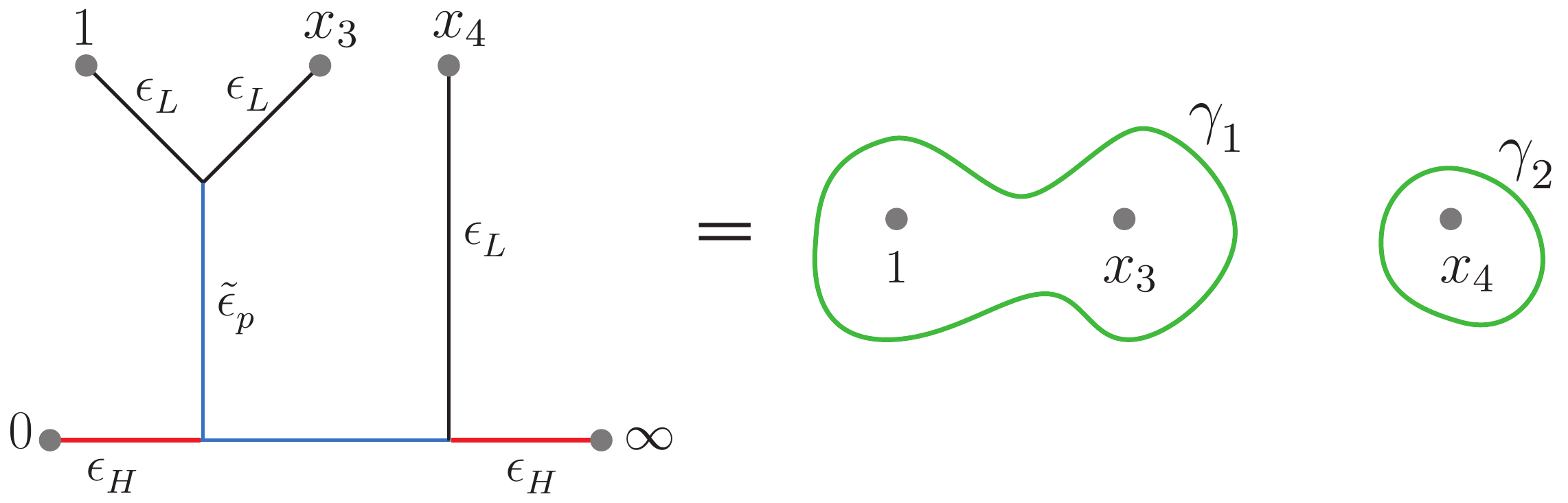} 
	\end{tabular}
	\caption[]{\small OPE channel and monodromy contours for $\Om_1$ for the 5-point block.}	\label{3-1}
\end{figure}

We are now presented with the task of solving \eqref{51} for the accessory parameters $c_3$ and $c_4$.  
This is the point where one can utilize the method of seed solutions introduced  in \cite{Alkalaev:2015lca}. The idea behind this method is that, in order to solve the monodromy problem for the conformal block having $m$ light operators, accessory parameters can be inherited from a lower point conformal block having $(m-1)$ light operators. More precisely, for the block with $m$ light insertions one uses the accessory parameters of the block with $(m-1)$ light insertions as zeroth-order solutions and then deforms these by the light parameter $\ep_L$.  

 Following \cite{Alkalaev:2015lca}, we choose a seed solution for the accessory parameter $c_3$ to be  the same of that of the 4-point conformal block \cite{FKW}. 
\begin{align}\label{ccc3}
c_3 = \frac{-\ep_L (\alpha-1+x_3^\alpha(\alpha+1))+\tep_p x_3^{\alpha/2}\alpha}{x_3(x_3^\alpha-1)} + \mathcal{O}(\ep_L^2 ) .
\end{align}
(Since, we shall be working the heavy-light limit -- i.e.~$c^\zero_i$ in \eqref{pert-sol} -- we shall drop the $\mathcal{O}(\ep_L^2 )$ from now on for brevity.)
Substituting this in \eqref{51} we get
\begin{align}\label{ccc4}
c_4 = - \frac{\ep_L}{x_4} .
\end{align}
The integrability condition \eqref{integrability-01} is trivially satisfied.

It is crucial to note that in \cite{Alkalaev:2015lca} an additional expansion was considered in their parameter $\ep_3$ corresponding to one of the light operators\footnote{See  equation (3.5) of \cite{Alkalaev:2015lca}.}. In our case, all light parameters have the same conformal dimension $\ep_L$. One can still consider a correction term of the form $\ep_L c^{\text{(corr)}}_3$ to $c_3$ above. However, it can be explicitly seen  that $c^{\text{(corr)}}_3=0$ because \eqref{ccc3} and \eqref{ccc4} are indeed the solutions to linear order in $\ep_L$ for the monodromy condition \eqref{51}. Hence, unlike \cite{Alkalaev:2015lca},  there are no additional corrections to \eqref{ccc3} and the seed solution for $c_3$ fully captures the accessory parameter in the heavy-light limit. This simplifying feature is special to the OPE channels or the corresponding monodromy contours we have considered and is also due the fact that all the light operators here have the same conformal dimension.


We can now use \eqref{acc-para} to obtain the conformal block. Upon integrating the accessory parameters, we get
\begin{align}
 f_\funf(x_3,x_4;\ep_L,\ep_H;\ep_p)=& \left[  \ep_L \left((1-\alpha)\log x_3 + 2 \log \frac{1-x_3^\alpha}{\alpha}\right)   +2 \tilde\ep_p\log\left[{4\alpha}\frac{1+x_3^{\alpha/2}}{1-x_3^{\alpha/2}}\right] \right] 
 \nn \\ & \qquad
+ \ep_L \log x_4\nn \\ 
=& \ f_\fear (1,x_3;\ep_L,\ep_H;\tilde\ep_p) +  \ep_L \log x_4.
\end{align}
The integration constants are chosen to be such that $f_\funf \sim (2\ep_L-\tep_p)\log (1-x_3)$ for $x_3 \to 1$ \cite{FKW,Hijano:2015rla}\footnote{This follows from \eqref{break-3pt} and considering the behaviour of the 3-point function $\la  \ol(1)\ol(x_3) \widetilde{\so}_{p}(0)\ra$ in the limit $x_3 \to 1$. }.
Here, $f_\fear$ is the function appearing in  the exponential of the    conformal block ($\cF_\fear=\exp(-cf_\fear/6)$) of the 4-point function $\la \oh(\infty) \ol(x_i) \ol(x_j) \oh(0)  \ra$ \cite{FKW}. 
\begin{align}\label{four-point-block}
 f_\fear(x_i,x_j;\ep_L,\ep_H;\ep_p) =& \ \ep_L \left((1-\alpha)\log x_ix_j + 2 \log \frac{x_i^\alpha-x_j^\alpha}{\alpha}  \right) 
 +2 \tilde\ep_p\log\left[ {4\alpha}\frac{x_j^{\alpha/2}+x_i^{\alpha/2}}{x_j^{\alpha/2}-x_i^{\alpha/2}}\right].
\end{align}
We now use the exponentiation of the conformal block \eqref{block-exponentiation} to obtain
\begin{align}\label{fac5}
\mathcal{F}_\funf(1,x_3,x_4;\ep_L,\ep_H;\tep_p)_{\Om_1} = \ &\exp\left[-\frac{c}{6} f_\fear(1,x_3;\ep_L,\ep_H;\tep_p)\right]   \times  x_4^{-h_L} \nn \\
=  \ & \cF _\fear(1,x_3;\ep_L,\ep_H;\tep_p) \times  x_4^{-h_L} .
\end{align}
The first factor here is the conformal block of the 4-point function $\la\oh(\infty) \ol(1) \ol(x_3) \oh(0)\ra$. The second factor of $x_4^{-h_L}$ is due to the 3-point function of $\la\oh(\infty)  \ol(x_4) \oh(0)  \ra$ normalized by $\la\oh(\infty)    \oh(0)  \ra$. 
\def\bsig{\bar{\sigma}}

It is worthwhile noting that this factorization shown above is true only at the level of conformal blocks -- and not correlators -- for a generic $\ol$. This is because not much is known about precise spectrum at low conformal dimension. In order to make a rigorous statement on the correlation function, one would require information about the structure constants $c_{LLa}$.

%
%

However, for the light operators being twist operators $\sigma_n,\,\bsig_n$, in the limit $n \to 1$ (relevant for entanglement entropy) the factorization \eqref{fac5}  shown above can  be independently seen (at the level of correlation functions) in terms of the OPE in the regime $x_3 \to 1$. This was previously shown in \cite{Watanabe}.
 It is known that twist operators fuse into the identity \cite{Calabrese:2004eu,Lunin:2000yv}\footnote{We shall be  implicitly working in the $n \to 1$ limit to ensure the operators are light.}
\begin{align}\label{ll-ope}
\sigma_n(1)\bsig_n(x_3)  \sim \mathbb{I} + \cO((1-x_3)^ r)\ ,  \qquad r \in \mathbb{Z}^+
\end{align}
This means that $\tep_p$ is 0 in the intermediate channel after fusion of two light operators. Moreover, the only operators which will arise in the intermediate channel are the identity and its descendants -- this includes the stress tensor and operators made of the derivatives and powers of the stress-tensor \cite{Hartman2}.  

Let us insert a complete set of states in the 5-point function and make use of the OPE \eqref{ll-ope}
\begin{align}\label{watanabe1}
&\la\oh(\infty) \sigma_n(1) \bsig_n(x_3) \sig_n(x_4) \oh(0)\ra_{x_3 \to 1} \nn \\
&=   \sum_{\alpha} \la\oh(\infty)\sigma_n(1) \bsig_n(x_3)\ket{\alpha}\bra{\alpha} \sig_n(x_4) \oh(0)\ra .
\end{align}
Since the leading term in the OPE \eqref{ll-ope} is the identity operator, the only non-zero contribution will arise from $\ket{\alpha}=\ket{\oh}$, due to orthonormality of states. Hence, just a single term in the above sum contributes and we have the factorization 
\begin{align}\label{watanabe2}
&\la\oh(\infty)\sigma_n(1)\bsig_n(x_3)  \sigma_n(x_4) \oh(0)\ra_{x_3 \to 1} \nn \\
&=    \la\oh(\infty)\sigma_n(1)\bsig_n(x_3)   \oh(0)\ra \la \oh(\infty) \sigma_n(x_4) \oh(0)\ra .
\end{align}
The 4-point function above will receive contributions solely from the identity block owing to \eqref{ll-ope}. 

\subsubsection*{$\Om_2$ channel}
The contour configuration for this case is equivalent to that of $\Om_1$ upon the replacements $x_3 \leftrightarrow x_4$ and $c_3 \leftrightarrow c_4$. The analysis for this channel is therefore exactly the same as that of $\Om_1$ with these exchanges. 
The final result for the conformal block in this channel is 
\begin{align}
\mathcal{F}_\funf(1,x_3,x_4;\ep_L,\ep_H;\tep_p)_{\Om_2} = \cF (1,x_4;\ep_L,\ep_H;\tep_p) \times  x_3^{-h_L} 
\end{align}

\def\cX{\mathcal{X}}
\subsubsection*{$\Om_3$ channel}
 
\begin{figure}[!t]
	\centering
	\begin{tabular}{c}
		\includegraphics[width=.75\textwidth]{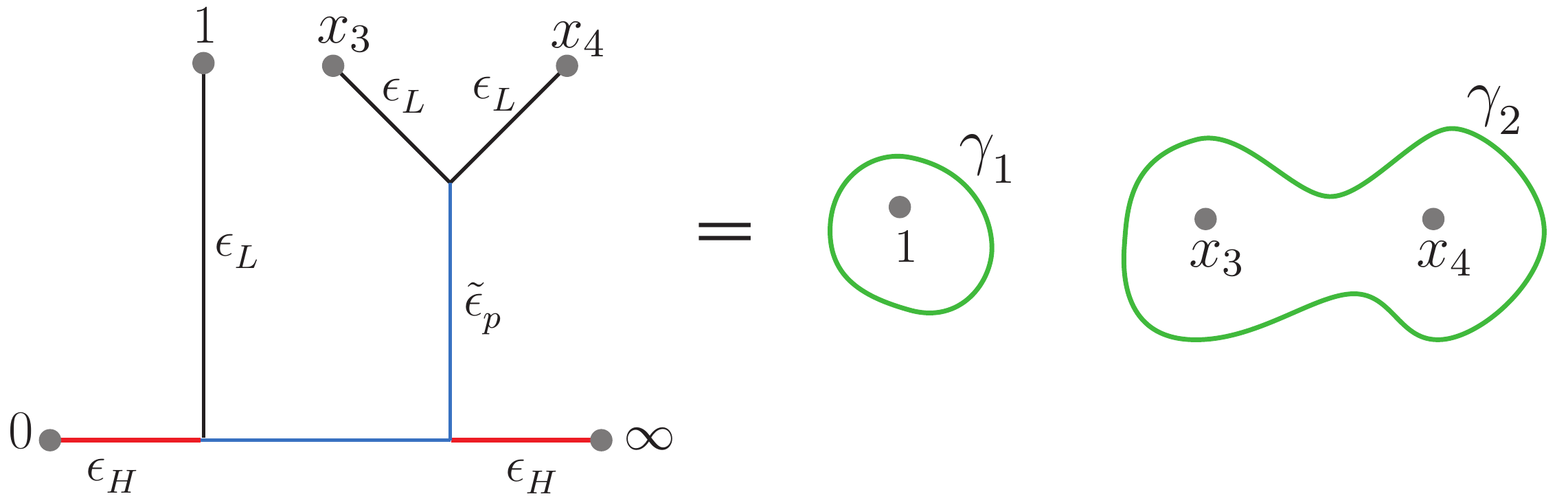} 
	\end{tabular}
	\caption{\small OPE channel and monodromy contours for $\Om_3$ for the 5-point block.}	\label{3-3}
\end{figure}
This is the OPE channel which considers fusion of the light primaries $\ol(x_3)$ and $\ol(x_4)$. The monodromy contours are shown in Fig.\,\ref{3-3}. 
The monodromy conditions \eqref{eigen-condition} for the two contours are
\begin{align}\label{52}
\cX[\gamma_1]=-4 \pi^2 \ep_L^2  = -4 \pi^2 \tilde\ep_p^2 
\end{align}
and 
\begin{align} \label{53}
\mathcal{X}[\gamma_2]=- &\frac{4\pi^2  x_3^{-\alpha/2}x_4^{-\alpha/2}}{\alpha^2}[\ep_L((\alpha-1)x_4^\alpha+(\alpha+1)x_3^\alpha)+(x_3^\alpha-x_4^\alpha) x_3 c_3]    \nn \\ &\quad\times [\ep_L((\alpha-1)x_3^\alpha+(\alpha+1)x_4^\alpha)+(x_4^\alpha-x_3^\alpha) x_4 c_4]   \  = \ - 4\pi^2 \tilde\ep_q^2   .
\end{align}
\eqref{52} gives $\ep_L=\tep_q$.
Inspired by the accessory parameters of the $\Om_1$ and $\Om_2$ channels, one can make an educated guess for the seed solution $c_3$
\begin{align}
c_3 = \frac{-\ep_L (x_4^\alpha(\alpha-1)+x_3^\alpha(\alpha+1)+\tep_p x_3^{\alpha/2}x_4^{\alpha/2}}{x_3(x_3^\alpha-x_4^\alpha)} .
\end{align}
Substituting this in \eqref{53} we obtain 
\begin{align}
c_4 = \frac{-\ep_L (x_3^\alpha(\alpha-1)+x_4^\alpha(\alpha+1)+\tep_px_4^{\alpha/2}x_3^{\alpha/2}}{x_4(x_4^\alpha-x_3^\alpha)}.
\end{align}
The integrability condition \eqref{integrability-01} is non-trivally satisfied in this case. Integrating the accessory parameters, we obtain
\begin{align}
f_\funf(1,x_3,x_4;\ep_L,\ep_H;\tep_p) = \ &\ep_L \left((1-\alpha)\log x_3x_4 + 2 \log \frac{x_3^\alpha-x_4^\alpha}{\alpha}  \right) 
+2 \tilde\ep_p\log\left[\frac{1}{4\alpha}\frac{x_4^{\alpha/2}+x_3^{\alpha/2}}{x_4^{\alpha/2}-x_3^{\alpha/2}}\right].
\end{align}
This result can also be written in terms of $f_\fear$ appearing in the 4-point conformal block \eqref{four-point-block}
\begin{align}
f_\funf(1,x_3,x_4;\ep_L,\ep_H;\tep_p) = f_\fear(1,x_3,x_4;\ep_L,\ep_H;\tep_p) + \ep_L \log 1 
\end{align}
The second term is obviously zero but we have retained it to preserve the structure we found in the previous channels. Exponentiating the above expression using \eqref{block-exponentiation}, we get
\begin{align}
\mathcal{F}_\funf(1,x_3,x_4;\ep_L,\ep_H;\tep_p)_{\Om_3} =\ &\cF (x_3,x_4;\ep_L,\ep_H;\tep_p) \times  (1)^{-h_L} .
\end{align}
which is the same factorization observed in the channels $\Om_1$ and $\Om_2$. 


\subsection*{\textit{Warmup example II : 6-point function}}

Our second example is  the conformal block of the 6-point function
\begin{align}
\la \oh(\infty) \ol(1) \ol(x_3) \ol(x_4) \ol(x_5) \oh(0) \ra .
\end{align}
This is the case of $m=4$. 

\subsubsection*{Perturbative expansion of the monodromy equation}

Once again, we start with the perturbative expansion in the parameter $\ep_L$ for $\psi(z)$, $T(z)$ and $c_i(z)$ \eqref{pert-sol}. The zeroth and first order equations are the same as that of \eqref{mono-0} and \eqref{mono-1}. 
The stress-tensor at the first two orders is
\begin{align}
T^{(0)} (z) =\ &\frac{\ep_H}{z^2}, \\
T^{(1)} (z) = \ & \frac{\ep_L}{(z-1)^2} + \frac{\ep_L}{(z-x_3)^2} +  \frac{\ep_L}{(z-x_4)^2}+ \frac{\ep_L}{(z-x_5)^2} - \frac{4 \ep_L}{z(z-1)}  \\
& \quad + \frac{x_3(1-x_3)}{z(z-1)(z-x_3)} c^{(0)}_3+ \frac{x_4(1-x_4)}{z(z-1)(z-x_4)} c^{(0)}_4+ \frac{x_5(1-x_5)}{z(z-1)(z-x_5)} c^{(0)}_5 \nn .
\end{align}
The solution to the zeroth order ODE in \eqref{mono-0} remains the same
\begin{align}
\psi_\pm^{(0)} (z) = z^{ (1\pm \alpha)/2} , \quad \quad \alpha = \sqrt{1-4 \ep_H}.
\end{align}
Similar to the 5-point case, we proceed to study monodromy constraints of the first order solution $\psi^\ein(z)$ which in turn is obtained by the method of variation of parameters \eqref{mvp}. 
\subsubsection*{Monodromy conditions}

\begin{figure}[!t]
	\centering
	\begin{tabular}{c}
		\includegraphics[width=.99999\textwidth]{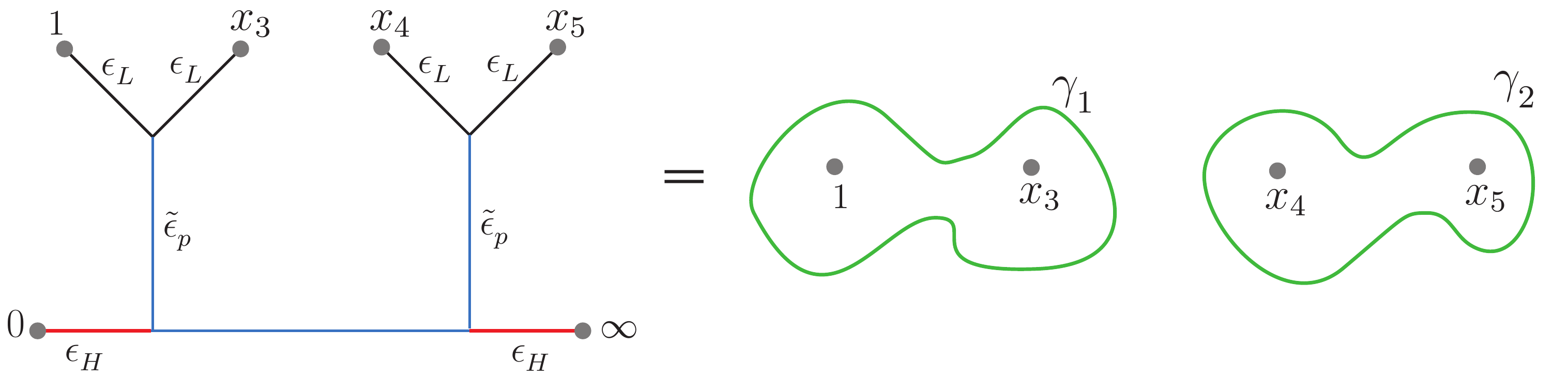} 
		\end{tabular}
		\caption{\small The OPE channel and contour configuration of $\Om_1$ for the case of the 6-point block.}
		\label{6-1}
		\end{figure}
There are three choices of contours in this case
\begin{itemize}
   \item $\Om _1$ : $\gamma_1$ enclosing $\lb 1,x_3 \rb$ and $\gamma_2$ enclosing $\lb x_4,x_5 \rb$
    \item  $\Om _2$ : $\gamma_1$ enclosing $\lb 1,x_5 \rb$ and $\gamma_2$ enclosing $\lb x_3,x_4 \rb$
     \item  $\Om _3$ : $\gamma_1$ enclosing  $\lb 1,x_4 \rb$ and $\gamma_2$ enclosing $\lb x_3,x_5 \rb$
\end{itemize}
The above channels are often referred to as $s$, $t$ and $u$ respectively. 
The monodromy contour and the corresponding OPE channel for $\Om_1$ are shown in the Fig.\,\ref{6-1}.

The monodromy matrix for the contours $\gamma_k$ above upto first order in $\epsilon_L$ has the same form as the one which we encountered in the 5-point block case \eqref{mono-matrix-01} with the elements given by the contour integrals in \eqref{mono-elements}. Furthermore, we impose the same monodromy condition $\cX[\gamma_i]$ for each of the contours in the configurations above \eqref{eigen-condition}. 

\subsubsection*{$\Om_1$ channel}
For the contours of the  $\Om_1$ channel we have 
\begin{align}\label{s1}
\cX [\gamma_1] =& -\frac{4\pi^2}{\alpha^2}x_3^{-\alpha} \left[   \ep_L (\alpha-1+x_3^\alpha(\alpha+1)) + c_3 x_3 (x_3^\alpha-1) \right]\nn \\
&\quad  \times \left[ \ep_L (\alpha-3 + x_3^\alpha(\alpha+3)+(x_3^\alpha-1)(c_3x_3+c_4x_4+c_5x_5))   				\right] = -4\pi^2 \tep_p^2 , \\ \label{s2}
\cX [\gamma_2] =& - \frac{4\pi^2 x_4^{-\alpha}x_5^{-\alpha}}{\alpha^2} \left[  \ep_L( (\alpha-1)  x_5^\alpha + (\alpha+1) x_4^\alpha) + x_4 (x_4^\alpha-x_5^\alpha)c_4 \right] \nn \\
&\quad  \times\left[ \ep_L (\alpha-1)  x_4^\alpha + (\alpha+1) x_5^\alpha)  - c_5x_5 (x_4^\alpha-x_5^\alpha)   \right]\ = \ -4\pi^2 \tep_q^2 . 
\end{align}
We shall focus on the case when the conformal dimensions of the operators in the  channels right after fusion with the light primaries are the same. That is, $\tep_p=\tep_q$. This is a major simplification, which was also used in \cite{Alkalaev:2015lca}, to facilitate the analysis.
We use the result of the accessory parameter $c_3$ from the 4-point (or 5-point) conformal block  as the  seed solution. As we had seen in the previous subsection, this seed solution is 
\begin{align}
c_3 = \frac{-\ep_L (\alpha-1+x_3^\alpha(\alpha+1)) + x_3^{\alpha/2}\alpha\tep_p}{x_3(x_3^\alpha-1)} . 
\end{align}
Substituting this into the monodromy conditions \eqref{s1} and \eqref{s2} we have two simultaneous equations in $c_4$ and $c_5$ which can be solved
\begin{align}
c_4 &=  \frac{-\ep_L (x_5^\alpha(\alpha-1)+x_4^\alpha(\alpha +1 )) + x_4^{\alpha/2}x_5^{\alpha/2} \alpha \tep_p }{x_4 (x_4^\alpha - x_5 ^\alpha)} , \\
c_5 &= \frac{-\ep_L (x_4^\alpha(\alpha-1)+x_5^\alpha(\alpha +1 )) + x_4^{\alpha/2}x_5^{\alpha/2} \alpha \tep_p }{x_5 (x_5^\alpha - x_4 ^\alpha)} .
\end{align}
It can be verified that the integrability condition \eqref{integrability-01} is satisfied.


We can now use \eqref{acc-para} to obtain the conformal block. 
\begin{align}
& f_\sechs(x_3,x_4,x_5;\ep_L,\ep_H;\tep_p)\nn \\ &=  \left[  \ep_L \left((1-\alpha)\log x_3 + 2 \log \frac{1-x_3^\alpha}{\alpha}\right)   + 2 \tilde\ep_p\log\left[ {4\alpha}\frac{1+x_3^{\alpha/2}}{1-x_3^{\alpha/2}}  \right] \right]    \\
&\qquad +\left[  \ep_L \left((1-\alpha)\log x_4x_5 + 2 \log \frac{x_4^\alpha-x_5^\alpha}{\alpha}  \right)  +2 \tilde\ep_p\log\left[ {4\alpha}\frac{x_4^{\alpha/2}+x_5^{\alpha/2}}{x_4^{\alpha/2}-x_5^{\alpha/2}}\right]\right]  . \nn
\end{align}
The structure of each of the terms in square brackets is yet again of the same form as that of 4-point conformal block case \eqref{four-point-block}. 
We now use the exponentiation of the conformal block \eqref{block-exponentiation}, to obtain
\begin{align}
\mathcal{F}_\sechs(1,x_3,x_4,x_5;\ep_L,\ep_H;\tep_p)_{\Om_1} = \ &\exp\left[-\frac{c}{6} f_\fear(1,x_3;\ep_L,\ep_H;\tep_p )\right]  
\times  \exp\left[-\frac{c}{6}f_\fear(x_4,x_5;\ep_L,\ep_H;\tep_q )\right] \nn \\
= \ & \mathcal{F}_\fear(1,x_3 ;\ep_L,\ep_H;\tep_p) \mathcal{F}_\fear(x_4,x_5 ;\ep_L,\ep_H;\tep_q) .
\end{align}
The 6-point conformal block, therefore, factorizes into a product of two 4-point ones. 

\def\twist{\sig_n}
\def\atwist{\bsig _n}

{Note that this factorization is not due to a decoupling of equations involving accessory parameters and is therefore not a mere doubling of the reduced problem for a single conformal block. In particular, equation \eqref{s1} contains all the accessory parameters and it is not a priori obvious from the monodromy method that this factorization will happen.}

Just like the case of the 5-point conformal block this factorization can be anticipated for the special case of twist operators (in the limit $n\to 1$). Inserting a complete set of states in the 6-point function and using the OPE channels  \eqref{ll-ope} relevant in the regimes $x_3 \to 1 $ or $ x_4 \to x_5  $  we have
\begin{align}
&\la\oh(\infty) \twist(1) \atwist(x_3) \twist(x_4) \atwist(x_5) \oh(0)\ra _{x_3 \to 1 \text{ and/or } x_4 \to x_5  } \nn \\
&=   \sum_{\alpha} \la\oh(\infty)  \twist(1) \atwist(x_3) \ket{\alpha}\bra{\alpha} \twist(x_4) \atwist(x_5)  \oh(0)\ra .
\end{align}  
Due to orthonomality of the complete set of states inserted, the only contribution will arise from $\ket{\alpha}=\ket{\oh}$, leading to the factorization 
\begin{align}
&\la\oh(\infty) \twist(1) \atwist(x_3) \twist(x_4) \atwist(x_5) \oh(0)\ra _{x_3 \to 1 \text{ and/or } x_4 \to x_5  }\nn \\
&=    \la\oh(\infty) \twist(1) \atwist(x_3)  \oh(0)\ra \la \oh(\infty)  \twist(x_4) \atwist(x_5)  \oh(0)\ra .
\end{align}
 
\subsubsection*{$\Om_2$ and $\Om_3$ channels}
The analysis in the $\Om_2$ and $\Om_3$ channels   proceeds exactly in the same manner as that of the $\Om_1$ channel with the replacements $x_3 \leftrightarrow x_4$ and  $x_3 \leftrightarrow x_5$ respectively. We obtain the factorizations 
\begin{align}
\mathcal{F}_\sechs(1,x_3,x_4,x_5;\ep_L,\ep_H;\tep_p)_{\Om_2}  
= \ & \mathcal{F}_\fear(1,x_4 ;\ep_L,\ep_H;\tep_p) \mathcal{F}_\fear( x_3,x_5 ;\ep_L,\ep_H;\tep_q) \\
\mathcal{F}_\sechs(1,x_3,x_4,x_5;\ep_L,\ep_H;\tep_p)_{\Om_3}  
= \ & \mathcal{F}_\fear(1,x_5;\ep_L,\ep_H;\tep_p) \mathcal{F}_\fear( x_4,x_3 ;\ep_L,\ep_H;\tep_q) 
\end{align}

\subsection{Conformal block for an arbitrary number of light operator insertions}
 Equipped with the examples considered above, we now come to the discussion of the correlator 
 \begin{align}
 \Big\langle \oh(\infty) \left[ \ol(1) \prod_{i=3}^{m +1}\ol (x_i) \right] \oh(0) \Big\rangle  
 \end{align}
 which has an arbitrary (even or odd) number of light operator insertions. 
 
\subsubsection*{Perturbative expansion of the monodromy equation}
We are interested in studying the monodromy properties of the ODE \eqref{monodromy-01} with $T(z)$ given in \eqref{Tz-full}. 
Just like the warmup examples considered above, we start with the perturbative expansion in the parameter $\ep_L$ for the quantities $\psi(z)$, $T(z)$ and $c_i(z)$ \eqref{pert-sol}. The zeroth and first order equations are the same as that of \eqref{mono-0} and \eqref{mono-1}. 
The stress-tensor at the first two orders is
\begin{align}
T^{(0)} (z) \ = \ &\frac{\ep_H}{z^2} \\
T^{(1)} (z) \ = \ & \frac{\ep_L}{(z-1)^2} + \sum_{i=3}^{m+1}\frac{\ep_L}{(z-x_i)^2}  - \frac{m\, \ep_L}{z(z-1)} + \sum_{i=3}^{m+1} \frac{x_i(1-x_i)}{z(z-1)(z-x_i)} c_i   
\end{align}
The solution to the zeroth order ODE in \eqref{mono-0} is 
\begin{align}
\psi_\pm^{(0)} (z) = z^{ (1\pm \alpha)/2} , \quad \quad \alpha = \sqrt{1-4 \ep_H}
\end{align}
We now consider the monodromy constraints of the first order solution $\psi^\ein(z)$ which is obtained by the method of variation of parameters \eqref{mvp}. 

\subsubsection*{Contour configurations and their enumerations}
The analysis of the 5- and 6-point conformal blocks posits that the choice of contour configurations  is slightly different for even or odd $m$.

For the case of an even number of light operator insertions, we can form  contours containing a pair of points each in $\nu^{\text{(even)}}_m = m!/(2^{m/2}(m/2)!)$ ways. Each contour contains two light operators located within. The contours fall under two major classes. 
\begin{itemize}
\item $\gamma_{(1,r)}$ : containing the points 1 and $x_r$ with $r\geq 3$ .
\item $\gamma_{(p,q)}$ : containing the points $x_p$ and $x_q$ with $p \neq q$ and $p,q\geq  3$ .
\end{itemize}
We use $\Om_{i}$ as a label for the $i$th contour configuration which includes information of all the contours $\gamma^{(i)}_k$
\begin{align}
\Om_{i}= \bigcup_{\lb (p,q) \rb } \gamma_{(p,q)}^i \ .
\end{align}
See \fig{8-1} for an example involving 6 light operators and 2 heavy operators. 
\begin{figure}[!t]
	\centering
	\begin{tabular}{c}
		\includegraphics[width=\textwidth]{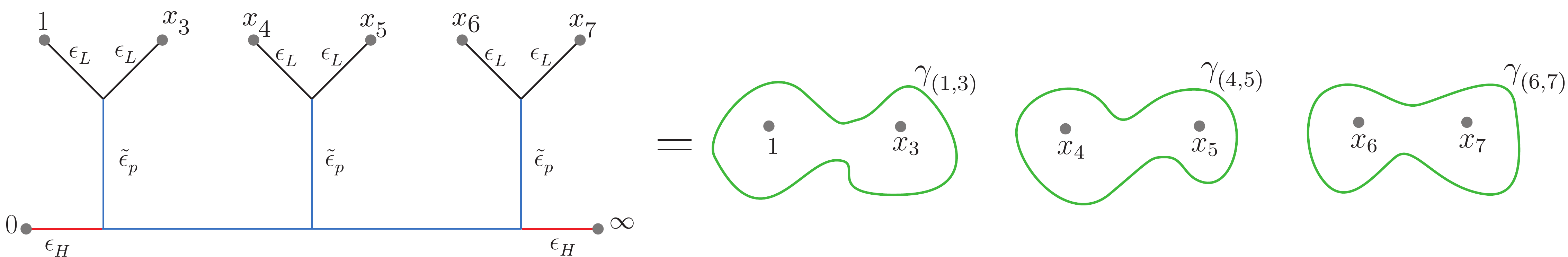} 
	\end{tabular}
	\caption{\small OPE channel and monodromy contours for $\Om_1$ for the 8-point block.}	\label{8-1}
\end{figure}

When an odd number of light operators are present there are $\nu^{\text{(odd)}}_m= $ $m!/(2^{(m-1)/2}$ $((m-1)/2)!)$ such contour configurations. For each contour configuration, there is a single contour containing just one light operator. All the other contours enclose a pair of light operators. The classes of contours in this case are four.
\begin{itemize}
	\item $\gamma_{(1)}$ : containing the point 1
	\item $\gamma_{(s)}$ : containing the point $x_s$ with $s\geq 3$
	\item $\gamma_{(1,r)}$ : containing the points 1 and $x_r$
	\item $\gamma_{(p,q)}$ : containing the points $x_p$ and $x_q$ with $p \neq q$ and $p,q\geq 3$ 
\end{itemize}
The set of contour configurations are of two types 
\begin{align}\label{odd1}
\Om^A_{i}= \gamma_{(r)} \cup \bigcup_{\lb (p,q) \rb} \gamma_{(p,q)}^i \quad
\text{and} \quad \Om^B_{i}= \gamma_{(1)} \cup \bigcup_{\lb (p,q) \rb } \gamma_{(p,q)}^i  \ .
\end{align}
An example of the $A$-type contour configuration is $\Om_1$ for the 5-point case shown in \fig{3-1} whereas the $\Om_3$ shown in \fig{3-3} is a $B$-type configuration. 

The contour configuration for the odd case ($m=2j+1$) is equivalent to that of the even case ($m=2j$) with an additional contour enclosing the extra point. Moreover, it can be seen that $\nu^{\text{(even)}}_{2j}=\nu^{\text{(odd)}}_{2j-1}$. (This number is 3 for both the 5- and 6-point cases.)

It can be seen that there is a one-to-one mapping between the contour configurations $\Om_i$ and the paired fusions of the light operators whose OPEs we consider to calculate the conformal block. We label this set of pairs as $\lb (p,q)\rb$. Often, we shall denote this map between the contour configurations and the set of fusions as
\begin{align*}
\Om_i \mapsto \lb (p,q)\rb  \ .
\end{align*}

\subsubsection*{Monodromy conditions}
The monodromy conditions are the same as the ones which we had imposed for the 5- and 6-point blocks. The final constraint is $\cX[\gamma_i]$ in equation \eqref{eigen-condition}. We shall now consider the even and odd number of  insertions of light operators separately. Additionally, we shall perform the analysis for the specific case of the conformal block in which the resulting intermediate channels from all pairwise fusions of light operators give the operators with the same conformal dimension. 
 
 \subsubsection*{\textit{Case I : Even number of light operator insertions}}

 Consider a specific contour configuration $\Om_i$. 
The monodromy condition for a contour enclosing $1$ and $x_r$ with $r=3,4,5,\cdots,m$ is
\begin{align}\label{one-r}
\mathcal{X}[\gamma_{(1,r)}] =& - \frac{4\pi^2}{\alpha^2} x_r^{-\alpha}\left[	\ep_L ((\alpha-1)+(\alpha+1)x_r^{\alpha})	+ (x_r^{\alpha} - 1)c_r x_r	\right] \\ 
&\times \left[ \ep_L ((\alpha-m+1)+(\alpha+m-1)x_r^{\alpha})	+(x_r^{\alpha} - 1) \sum_{i=3}^{m+1}c_i x_i			\right]  \ = \ - 4 \pi^2 \tep_a \nn .
\end{align}
and that for a contour enclosing $x_p$ and $x_q$ is
\begin{align} \label{pqq}
\mathcal{X}[\gamma_{(p,q)}] =& - \frac{4\pi^2}{\alpha^2} (x_px_q)^{-\alpha} \left[	\ep_L ((\alpha-1)x_q^{\alpha}+(\alpha+1)x_p^{\alpha})	+ (x_p^{\alpha} - x_q^{-\alpha})c_p x_p	\right] \\ 
&\qquad\qquad \times   \left[	\ep_L ((\alpha-1)x_p^{\alpha}+(\alpha+1)x_q^{\alpha})	+ (x_q^{\alpha} - x_p^{-\alpha})c_q x_q	\right]   \ = \ - 4 \pi^2 \tep_b .
 \nn 
\end{align}
As mentioned above, we shall set $\tep_a=\tep_b=\tep_p$.

Unlike the case of conformal blocks involving only an even number $(m)$ of light operators  considered in \cite{Hartman1}, the system of monodromy conditions with the additional heavy operators does not decouple into $m/2$ independent monodromy problems. It can be seen that the monodromy condition \eqref{one-r} involves all the accessory parameters. 

The analysis of the 6-point conformal block strongly suggests a guess for the solution of the above coupled equations
\begin{align}\label{r}
c_r 
&= \frac{-\ep_L (\alpha-1+x_r^\alpha(\alpha+1))+x_r^{\alpha/2}\alpha\tep_p}{x_r(x_r^\alpha-1)} , \\ \label{p}
c_p &= \frac{- \ep_L (x_q^\alpha (\alpha-1)+x_p^\alpha(\alpha+1))+(x_p x_q)^{\alpha/2}\alpha \tep_p}{x_p (x_p^\alpha-x_q^\alpha)} , \\   \label{q}
c_q&= \frac{- \ep_L (x_p^\alpha (\alpha-1)+x_q^\alpha(\alpha+1))+(x_q x_p)^{\alpha/2}\alpha\tep_p}{x_q (x_q^\alpha-x_p^\alpha)} . 
\end{align}
This obeys the integrability condition  $\pd_{x_p}c_q=\pd_{x_q}c_p$. Also, all other integrability conditions get trivially satisfied. 

Let us now verify this is indeed a solution to the coupled monodromy constraints \eqref{one-r} and \eqref{pqq}. It can be easily seen that $\cX[\gamma_{(p,q)}]$ is obeyed by $c_p$ and $c_q$ above. Consider the monodromy condition $\cX[\gamma_{(1,r)}]$ in \eqref{one-r} in which one of the contours contain $1$ and $x_r$. This monodromy condition involves all the accessory parameters. Note that, equations \eqref{p} and \eqref{q} give
\begin{align}\label{pq}
c_p x_p + c_q x_q = -2 \ep_L .
\end{align}
From \eqref{pq} and \eqref{r}, we have
\begin{align}
\sum_{i=3}^{m+1}  c_i x_i &= c_r x_r - (m-2)\ep_L \nn \\
&=\frac{ - \ep_L ((\alpha-m+1)+(\alpha+m-1)x_r^{\alpha}) + x_r^{\alpha/2} \alpha \tep_p}{(x_r^\alpha-1)} \label{sum-00} .
\end{align}
The first equality follows from \eqref{pq} and the fact that there are $(m-2)/2$ pairs of contours other than the one containing $1$ and $x_r$. Note that $r$ is fixed and there is no sum over $r$ in the first term $c_r x_r$.  It can then be very explicitly checked that substituting \eqref{sum-00} along with \eqref{r} in the LHS of equation \eqref{one-r} satisfies it giving $-4\pi^2 \tep_q^2$ (with $\tep_a=\tep_p$).

We can now use \eqref{acc-para} to obtain the conformal block by integrating the accessory parameters  \eqref{p},  \eqref{q} and \eqref{r}. The integration constants are fixed in the same manner as the warmup examples, by demanding the expected behaviour $(2\ep_L-\tep_p)\log(x_p-x_q)$ as $x_p \to x_q$.
\begin{align}\label{factor-formula}
&f(\lb x_i \rb;\ep_L,\ep_H;\tep_p )\ \nn \\ &= \ \sum_{\Om_i \mapsto {\lbrace(p,q)\rbrace}} \left(\ep_L(1-\alpha)\log x_px_q + 2 \ep_L\log \frac{x_p^\alpha-x_q^\alpha}{\alpha} +2 \tilde\ep_p\log\left[{4\alpha}\frac{x_p^{\alpha/2}+x_q^{\alpha/2}}{x_p^{\alpha/2}-x_q^{\alpha/2}}\right] \right)    \nn \\
&= \ \sum_{\Om_i \mapsto {\lbrace(p,q)\rbrace}} f_\fear(x_p,x_q;\ep_L,\ep_H;\tep_p).
\end{align}
Here the sum is over the set of $m/2$ contours containing a pair of light operators located at $x_p$ and $x_q$ respectively. This set also includes the contour containing $1$ and $x_r$. Equivalently the set $\lbrace(p,q)\rbrace$ is also the OPE channel along which we perform the conformal partial wave expansion.

Equation \eqref{factor-formula} upon exponentiation clearly shows the factorization of the $(m+2)$-point conformal block into $m/2$ 4-point blocks. 
\begin{align}\label{big-block}
\mathcal{F}_{(m+2)}(\lb x_i \rb ;\ep_L,\ep_H;\tep_p)=&\prod_{\Om_i\mapsto{\lbrace(p,q)\rbrace}}\exp\left[-\frac{c}{6}f_\fear(x_p,x_q;\ep_L,\ep_H;\tep_p)\right] \nn \\
=& \prod_{\Om_i\mapsto{\lbrace(p,q)\rbrace}} \cF_\fear (x_p,x_q;\ep_L,\ep_H;\tep_p).
\end{align}
This proves that the $(m+2)$-point conformal block, in the specific classes of  OPE channels analysed above, factorizes into 4-point conformal blocks. This is a central result of this paper.

\subsubsection*{\textit{Case II : Odd number of light operator insertions}}

As noted before, the contour configurations for an odd number of insertions, $m$,  is the  sum of the even number contour configurations for $m-1$ light operator insertions plus another contour enclosing a single point. 

For the contour configuration of the first kind \eqref{odd1}, the monodromy conditions are 
\begin{align}
\cX[\gamma_{(1,r)}] =& - \frac{4\pi^2}{\alpha^2} x_r^{-\alpha}\left[	\ep_L ((\alpha-1)+(\alpha+1)x_r^{\alpha})	+ (x_r^{\alpha} - 1)c_r x_r	\right] \\ 
&\quad \times \left[ \ep_L ((\alpha-m+1)+(\alpha+m-1)x_r^{\alpha})	+(x_r^{\alpha} - 1) \sum_{i=3}^{m+1}c_i x_i			\right]  \ = \ - 4 \pi^2 \tep_a \nn ,
\end{align}
and that for a contour enclosing $x_p$ and $x_q$ is
\begin{align} 
\mathcal{X}[\gamma_{(p,q)}] =& - \frac{4\pi^2}{\alpha^2} (x_px_q)^{-\alpha} \left[	\ep_L ((\alpha-1)x_q^{\alpha}+(\alpha+1)x_p^{\alpha})	+ (x_p^{\alpha} - x_q^{-\alpha})c_p x_p	\right] \\ 
&\qquad\qquad \times   \left[	\ep_L ((\alpha-1)x_p^{\alpha}+(\alpha+1)x_q^{\alpha})	+ (x_q^{\alpha} - x_p^{-\alpha})c_q x_q	\right]   \ = \ - 4 \pi^2 \tep_b 
\nn .
\end{align}
For the contours enclosing the single points, they are
\begin{align}
\cX[\gamma_{(r)}]=-4 \pi^2 \ep_L^2  &= -4 \pi^2 \tilde\ep_c^2 ,\\
\cX[\gamma_{(1)}]=-4 \pi^2 \ep_L^2  &= -4 \pi^2 \tilde\ep_d^2 .
\end{align}
$\cX[\gamma_{(s)}]$ and $\cX[\gamma_{(1)}]$ imply that the conformal dimension of the operator living within the contour is picked by the residue. 

Similar to the previous cases, we shall now set $\tep_a=\tep_b=\tep_p$. For the configuration of the first kind, $\Om^A_{i}$ in \eqref{odd1}, consisting of the contours $\gamma_{(r)}$, $\gamma_{(p,q)}$ and $\gamma_{(1,r)}$, the solutions for the accessory parameters are
\begin{align} 
c_r 
&= \frac{-\ep_L (\alpha-1+x_r^\alpha(\alpha+1))+x_r^{\alpha/2}\alpha\tep_p}{x_r(x_r^\alpha-1)} , \\  
c_p &= \frac{- \ep_L (x_q^\alpha (\alpha-1)+x_p^\alpha(\alpha+1))+(x_p x_q)^{\alpha/2}\alpha \tep_p}{x_p (x_p^\alpha-x_q^\alpha)} ,\\    
c_q&= \frac{- \ep_L (x_p^\alpha (\alpha-1)+x_q^\alpha(\alpha+1))+(x_q x_p)^{\alpha/2}\alpha\tep_p}{x_q (x_q^\alpha-x_p^\alpha)} ,\\
c_s&= - \frac{\ep_L}{x_s} .
\end{align}
The above accessory parameters can be integrated \eqref{acc-para} to obtain the function appearing in the exponential of the conformal block. 
\begin{align}\label{factor-formula-odd1}
&f(\lb x_i \rb;\ep_L,\ep_H;\tep_p)\ \nn \\ &= \ \ep_L \log x_s + \sum_{\Om^A_i\mapsto{\lbrace(p,q)\rbrace}} \left(\ep_L(1-\alpha)\log x_px_q + 2 \ep_L\log \frac{x_p^\alpha-x_q^\alpha}{\alpha} +2 \tilde\ep_p\log\left[{4\alpha}\frac{x_p^{\alpha/2}+x_q^{\alpha/2}}{x_p^{\alpha/2}-x_q^{\alpha/2}}\right] \right)    \nn \\
&= \ \ep_L \log x_s +\sum_{\Om^A_i\mapsto{\lbrace(p,q)\rbrace}} f_\fear(x_p,x_q;\ep_L,\ep_H;\tep_a).
\end{align}
This upon exponentiation gives 
\begin{align}\label{big-block-odd1}
\mathcal{F}_{(m+2)}(\lb x_i \rb ;\ep_L,\ep_H;\tep_p)=& \ (x_s)^{-\ep_L}\prod_{\Om^A_i\mapsto{\lbrace(p,q)\rbrace}}\exp\left[-\frac{c}{6}f_\fear(x_p,x_q;\ep_L,\ep_H;\tep_p)\right] \nn \\
= &  \ (x_s)^{-\ep_L} \prod_{\Om^A_i\mapsto{\lbrace(p,q)\rbrace}} \cF_\fear (x_p,x_q;\ep_L,\ep_H;\tep_p) .
\end{align}
This shows the factorization of the $(m+2)$-point block with odd $m$ into a 3-point function (without the structure constant) and $(m-1)/2$ number of 4-point conformal blocks. 

For the second kind of contour configuration, $\Om^B_{i}$  in \eqref{odd1},  consisting of $\gamma_{(1)}$ and $\gamma_{(p,q)}$  we have
\begin{align} 
c_p &= \frac{- \ep_L (x_q^\alpha (\alpha-1)+x_p^\alpha(\alpha+1))+(x_p x_q)^{\alpha/2}\alpha \tep_p}{x_p (x_p^\alpha-x_q^\alpha)} , \\    
c_q&= \frac{- \ep_L (x_p^\alpha (\alpha-1)+x_q^\alpha(\alpha+1))+(x_q x_p)^{\alpha/2}\alpha\tep_p}{x_q (x_q^\alpha-x_p^\alpha)} .
\end{align}
Integrating these, we obtain \eqref{acc-para}
\begin{align}\label{factor-formula-odd2}
&f(\lb x_i \rb;\ep_L,\ep_H;\tep_p)\ \nn \\ &= \ \ep_L \log 1 + \sum_{\Om^B_i\mapsto{\lbrace(p,q)\rbrace}} \left(\ep_L(1-\alpha)\log x_px_q + 2 \ep_L\log \frac{x_p^\alpha-x_q^\alpha}{\alpha} +2 \tilde\ep_p\log\left[{4\alpha}\frac{x_p^{\alpha/2}+x_q^{\alpha/2}}{x_p^{\alpha/2}-x_q^{\alpha/2}}\right] \right)    \nn \\
&= \ \ep_L \log 1 +\sum_{\Om^B_i\mapsto{\lbrace(p,q)\rbrace}} f_\fear(x_p,x_q;\ep_L,\ep_H;\tep_a) .
\end{align}
Exponentiating this to obtain the conformal block using \eqref{block-exponentiation}, we have 
\begin{align}\label{big-block-odd2}
\mathcal{F}_{(m+2)}(\lb x_i \rb ;\ep_L,\ep_H;\tep_a)=& \ (1)^{-\ep_L}\prod_{\Om^B_i\mapsto{\lbrace(p,q)\rbrace}}\exp\left[-\frac{c}{6}f_\fear(x_p,x_q;\ep_L,\ep_H;\tep_a)\right] \nn \\
= &  \ (1)^{-\ep_L} \prod_{\Om^B_i\mapsto{\lbrace(p,q)\rbrace}} \cF_\fear (x_p,x_q;\ep_L,\ep_H;\tep_a).
\end{align}
Hence, the factorization is also clear for the contour configuration, $\Om^B_{i}$. 

As we had seen the 5- and 6-point examples, the factorization for an arbitrary number of light operator insertions can  be expected for the light operators being twist and anti-twist operators. Since, twist operators fuse into the identity \eqref{ll-ope}, the correlator with $2N$ number of twist and anti-twist insertions will factorize into $N$ 4-point functions each having two heavy and two twists. This factorization will occur only in specific regimes in the space of $\lb x_i\rb$ where one can use the OPEs within  the correlator. 

There is one additional caveat to our monodromy analysis. 
A curious feature of the above conformal blocks is that they are apparently independent of the conformal dimensions of operators in other intermediate channels --- the horizontal channels $\tep_{Q,R,\cdots}$ in \fig{intro-fig} and \fig{int-heavy}. Let us evaluate   the conformal dimensions of the operators in these intermediate channels. 
Consider, $\tep_Q$ in \fig{int-heavy}. This can be obtained by evaluating the monodromy around the contour containing $\oh(0)$, $\ol(1)$ and $\ol(x_3)$ --- $\gamma_{(0,1,x_3)}$ in \fig{int-heavy}. At the leading order ($\ep_L^0$) the only contribution arises from $z=0$. This is equation \eqref{monodromyat0}. At the linear order in $\ep_L$, there   is a vanishing contribution from the residue at $z=0$ and the monodromy is effectively the same as that of the contour $\gamma_1$ in Fig.~\ref{3-1} and leads to equation \eqref{51}. Hence, from \eqref{monodromyat0} and \eqref{mono-matrix-01}, we have 
\begin{align}\label{horizontal}
\mathbb{M}(\gamma_{( 0,1,x_3)}) = -\begin{pmatrix}
e^{\pi i \alpha} &0 \\
0 &e^{-\pi i \alpha} \\
\end{pmatrix}  + \begin{pmatrix}
I^\kay_{++}({\gamma_2})  & I^\kay_{+-} ({\gamma_2})\\
I^\kay_{-+} ({\gamma_2}) & I^\kay_{--}({\gamma_2})
\end{pmatrix} ,
 \qquad \text{with, }\alpha=\sqrt{1-4\ep_H}.
\end{align}
Therefore, comparing \eqref{horizontal} with \eqref{mono-def} for the contour $\gamma_{(0,1,x_3)}$ which has $\Lambda=\sqrt{1-4\tep_Q}$, results in  $\tep_Q = \ep_H +\cO(\tep_p)$. Here, the $\mathcal{O}(\tep_p)$ term  arises from the second term in \eqref{horizontal} or equivalently from the contour $\gamma_1$ in Fig.~\ref{3-1}  as explained above.
 
   \begin{figure}[!t]
   	\centering
   	\begin{tabular}{c}
   		\includegraphics[width=.95\textwidth]{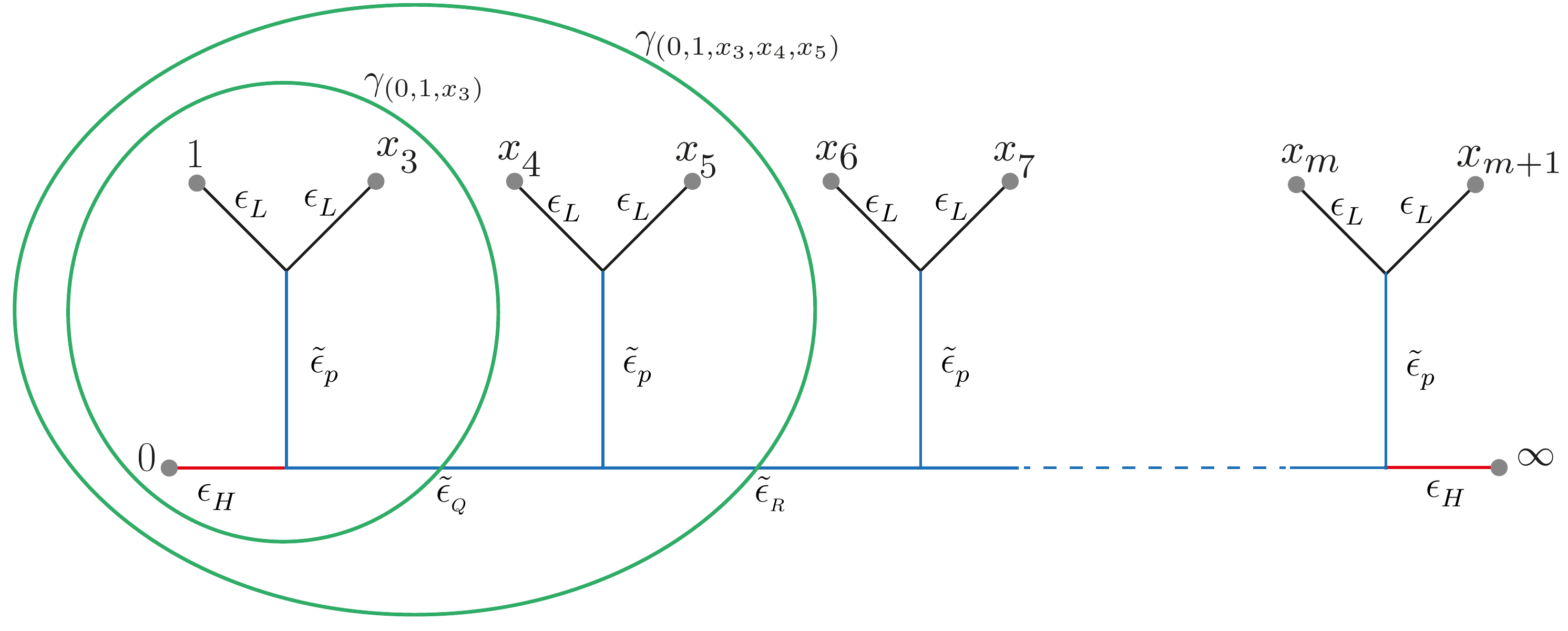} 
   	\end{tabular}
   	\caption{\small Monodromy contours to calculate $\tep_Q$ and $\tep_R$.}	\label{int-heavy}
   \end{figure}
	\red{We shall now make the only assumption on the spectrum and OPEs which was mentioned earlier in Section  \ref{sec:3}}. This is, the intermediate operator ($\tep_p$) appearing after fusion of two light operators has $\tep_p \ll \ep_L$ (and this automatically implies, $\tep_p \ll \ep_H$). 
	\red{Furthermore, we would be interested in the regime of holography in which the CFT's dual description in terms of a classical gravity approximation. The modes dual to the graviton fluctuations in the bulk corresponds those built of the stress-tensor (and its descendants) in the CFT. In other words, this is the Verma module of the vaccum which contain states (or equivalently operators) with conformal dimensions which are $\mathcal{O}(1)$ and do not scale as the central charge unlike the light and heavy operators and therefore satisfies $\tep_p \ll \ep_L$.  }
Hence, the dominant contribution to this monodromy \eqref{horizontal} comes from the heavy operator at $z=0$. 
This shows that in the heavy-light regime, this intermediate channel $\tep_Q$ is dominated by a heavy operator exchange $\ep_H$ \footnote{This fact is also supported by $\ket{\alpha}=\oh(0)\ket{0}$ in equations \eqref{watanabe1} and \eqref{watanabe2}. We shall also see the bulk counterpart of $\tep_p \ll \ep_L$ in Section \ref{sec:5} which was also previously used in \cite{Hijano:2015rla}.}.

In order to obtain $\tep_R$, one can repeat the above exercise by considering the monodromy of the contour containing $0$, 1, $x_3$, $x_4$ and $x_5$ --- $\gamma_{(0,1,x_3,x_4,x_5)}$ in \fig{int-heavy}. Once again, the dominant contribution will arise from the heavy operator at $z=0$, which leads to $\tep_R=\ep_H$. Continuing in this fashion, it can therefore be seen, that all the horizontal intermediate channels in \fig{int-heavy} are dominated by heavy operator exchanges in the heavy-light limit\footnote{It is reassuring  to observe that this also gives the same spectrum of eigenvalues for $ \mathbb{M}(\gamma_{( 0,1,x_3,\cdots,x_{m+1})})$ and $\mathbb{M}(\gamma_{( \infty)})$ -- as one should expect from the Riemann sphere.}. The dependence on the conformal dimension of these channels then enters the conformal block via the relation $\tep_Q=\tep_R=\cdots=\ep_H$. Therefore, the assumption, $\tep_p\ll \ep_L $ is necessary to have the heavy exchanges in the horizontal channels which results in the factorization of the higher-point block into 4-point blocks. 


It is worthwhile mentioning that there are other branches of solutions to the monodromy constraints for the accessory parameters. This point was emphasized in \cite{Alkalaev:2015lca}. It was shown in \cite{Alkalaev:2015lca} that only one of these branches matches with the one obtained from gravity. In our analysis above, we have restricted our attention solely to the branch which is relevant to make contact with holography in Section \ref{sec:5}. 


\def\Renyi{R\'{e}nyi }
\section{Entanglement entropy and mutual information of heavy states}\label{sec:4}

The results on conformal blocks obtained in the previous section can be utilized to evaluate the entanglement entropy of disjoint intervals in states excited by the heavy operator $\oh$. The single interval entanglement entropy of heavy states was considered in \cite{Hartman2}. Using the state-operator correspondence,  these `heavy states' can be obtained from the vacuum as\footnote{There is a slight abuse of notation here. The $\psi$ or $\Psi$ appearing in this section is neither the same nor related in any way  to $\psi$ or $\Psi$ which appeared in Section \ref{sec:3}.}
\[
\ket{\psi} = \oh(0) \ket{0} \quad \text{and} \quad \bra{\psi} =\lim\limits_{z,\bz \to \infty} \bz^{2h_H} z^{2h_H} \bra{0} \oh(z,\bz ).
\]

We briefly review the definitions of entanglement entropy and the replica trick used  to calculate it. The entanglement entropy is defined as the von-Neumann entropy corresponding to the reduced density matrix $\rho_\cA$
\begin{align}
S_\cA \ =  \ - \tr_\cA \, \rho_\cA \log \rho_\cA 
\end{align}
whilst the R\'{e}nyi entropy is obtained from the moments of $\rho_\cA$
\begin{align}
S^{(n)}_\cA = \frac{1}{1-n} \log \, \tr_\cA  \, (\rho_\cA )^n .
\end{align}
The reduced density matrix is, in turn, obtained by tracing out the Hilbert space lying outside $\cA$, i.e.~$\rho_\cA = \tr_{\cA'} \rho $. The full density matrix $\rho$ in our case in terms of the excited state is $\rho=\ket{\psi}\bra{\psi}$.
The \Renyi  entropies are a convenient computational tool, as it can be used to obtain the entanglement entropy by the analytic continuation to $n \to 1$. It can be shown via the path integral  \cite{Calabrese:2004eu} that the quantity $ \tr_\cA  \, (\rho_\cA )^n $ can be written in terms of the partition function of the $n$-sheeted Riemann surface (with each copy glued along $\cA$) as
\begin{align}
G_n \equiv  \tr_\cA  \, (\rho_\cA )^n  =  \frac{Z_n}{Z_1^n },
\end{align}
where, we have defined the normalized partition function $G_n$. 
The replica trick can be implemented by means of the twist operators, $\twist,\,\atwist$, which impose the required boundary conditions as one moves from one sheet to another. The conformal dimensions of the (anti-)twist operators are 
\begin{align}\label{twist-dim}
h_{\twist}=h_{\atwist} = \frac{c}{24} \left(n - \frac{1}{n}\right).
\end{align}
Hence, these operators become light in the limit relevant for entanglement entropy, $n\to 1$.

We shall focus on the case in which the sub-system $\cA$ is made of $N$ disjoint intervals i.e.~$\cA = \cup_i \cA_i$ \footnote{ It might be worth mentioning here, that the replica geometry for $N$ disjoint intervals is a surface of genus $(n-1)(N-1)$ from the Riemann-Hurwitz theorem \cite{Faulkner}. However, since we are interested in entanglement entropy (which is the $n \to 1$ limit of the \Renyi entropy, $S^{(n)}_\cA $) this is the limiting case of  genus-0 or a sphere. We are therefore allowed use the results for correlation functions on the plane. }. As shown in Fig.\,\ref{disjoint}, these intervals are located at $[1,x_3]$, $[x_4,x_5]$, \dots $[x_{2N},x_{2N+1}]$. In this setup, $x_i<x_j$ for all $i<j$. This ordering of the locations reduce the number of possible OPE channels. Also, the OPEs are non-vanishing only for a twist with an anti-twist operator and vanishing for a pair of twists (or a pair of anti-twists). The number of allowed OPE channels for $N$ disjoint intervals is actually, given by the Fibonacci number $F_{2N-1}$. (This is discussed further in Section \ref{sec:7}.)

\begin{figure}[!t]
	\centering
	\begin{tabular}{c}
		\includegraphics[width=.95\textwidth]{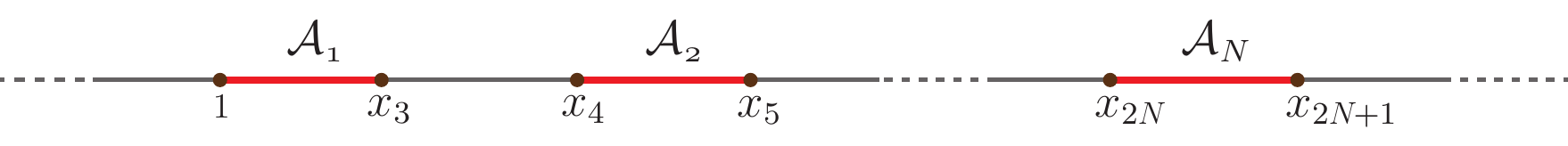} 
	\end{tabular}
	\caption{\small Configuration of disjoint intervals on a line.}	\label{disjoint}
\end{figure}
 For heavy states, the partition function on the $n$-sheeted Riemann surface is the following correlation function of twist operators 
\begin{align}\label{gn-corr}
G_n (x_i,\bx_i )= \bra{\Psi} \twist(1) \atwist(x_3) \twist(x_4) \atwist(x_5) \twist(x_6) \atwist(x_7) \dots \twist(x_{2N }) \atwist(x_{2N+1})  \ket{\Psi}.
\end{align}
Here, the state $\ket{\Psi}$ is the state in the $n$-sheeted replica which has insertions of $\oh$ throughout all its copies. That is, 
\begin{align}
\Psi = \prod_{i=1}^n (\oh)_i, \quad \text{and } h_{\Psi} = n h_H .
\end{align}
The correlator \eqref{gn-corr} can be rewritten as 
\begin{align}
G_n (x_i,\bx_i )= \bra{0} \ \Psi(\infty) \ \twist(1) \atwist(x_3)\  \prod_{i=4,6,\cdots}^{2N}\twist(x_i) \atwist(x_{i+1})  \ \Psi(0) \  \ket{0}.
\end{align}
One can evaluate this correlation function by decomposing into conformal blocks. As argued in \cite{Hartman2}, for a CFT at large central charge with a sparse spectrum of low-dimension operators, the dominant contribution to this correlator will arise from the identity block. We can therefore use the results derived in the previous section for the conformal block with an arbitrary even number of light operator insertions. However, it is important to remember that the number of OPE channels in this case will be reduced for reasons we mentioned earlier. We  denote these allowed OPE channels by $\tom_i$ (which is a subset of the channels $\Om_i$ in the case of even number of light insertions considered in the previous section). Thus, from equation \eqref{big-block}, we have, with $\tep_p=0$ for the identity block
\begin{align}\label{gn-f}
G_n(x_i,\bx_i)|_{n \to 1 }\approx&\ \mathcal{F}_{(2N+2)}(\lb x_i \rb ;\ep_L,\ep_H;0) \bar{\mathcal{F}}_{(2N+2)}(\lb \bx_i \rb ;\ep_L,\ep_H;0)\nn \\
=&\prod_{\tom_i \mapsto {\lbrace(p,q)\rbrace}}\exp\left[-\frac{nc}{6}f_\fear(x_p,x_q;\ep_L,\ep_H;0)\right]  \exp\left[-\frac{nc}{6}f_\fear(\bx_p,\bx_q;\ep_L,\ep_H;0)\right]  \nn \\
=&  \prod_{\tom_i \mapsto {\lbrace(p,q)\rbrace}} \cF_\fear (x_p,x_q;\ep_L,\ep_H;0)\ \bar{\cF}_\fear (\bx_p,\bx_q;\ep_L,\ep_H;0).
\end{align}
Note that the central charge above is $nc$ owing to $n$ replicas of the original theory. 
The essential object here is the function $f_\fear$ given by \eqref{four-point-block}. For this specific case, we have 
\begin{align}\label{ee-block}
 f_\fear(x_i,x_j;\ep_L,\ep_H;0) =& \  2 \ep_L   \log \frac{x_i^\alpha-x_j^\alpha}{\alpha (x_ix_j)^{\frac{\alpha-1}{2}}}  .
\end{align}
For the twist operators, $\ep_L = \frac{n^2-1}{4n}$, from \eqref{twist-dim}. 
Using the above relations in the limit $n \to 1 $, the entanglement entropy is given by 
\begin{align}\label{ee-disjoint}
S_\cA \ = \ \lim_{n\to 1 }\, S^{(n)}_\cA  
\ = \  \frac{c}{3} \min_{i}\Bigg\lb\sum_{\tom_i \mapsto \lbrace(p,q)\rbrace} \log  \frac{(x_p^\alpha-x_q^\alpha) }{\alpha  (x_px_q)^{\frac{\alpha-1}{2}}}  \Bigg\rb.
\end{align}
This is the final result for the entanglement entropy of $N$ disjoint intervals in the heavy state. The minimal condition above implies that one need to pick the relevant OPE channel ($\tom_i$) depending on the values of the cross-ratios $x_i$ (cf.~\cite{Hartman1}). The cross-ratios above are taken to be real ($x_i = \bx_i$) since in the Lorentzian picture the intervals are spacelike and one can consider them to be lying on the time slice $t=0$ without any loss of generality. The above result  is the excited state analogue to the one for vacuum derived in \cite{Hartman1,Headrick:2010zt}\footnote{This result was independently derived in \cite{Fitzpatrick:2015zha} in the lightcone OPE limit. This requires the light operators $\ol$ to be far from the heavy operators $\oh$.  }. 

For the case of two intervals, the mutual information can be straightforwardly calculated from \eqref{ee-disjoint}. Its definition is
\begin{align}
I_{\cA_i, \cA_j} =  S_{\cA_i} + S_{\cA_j}- S_{\cA_i \cup\cA_j}.
\end{align}
Without any loss of generality, we can choose the two intervals to be $[1,x_3]$ and $[x_4,x_5]$. From the channels $\Om_{1,3}$ for the 6-point block (consisting of 4 twists and 2 heavy operators) considered in subsection \ref{subsec:warmup}, we have
\begin{align}
I_{\cA_1,\cA_2} = \begin{dcases*}
 0						&\text{for $s$-channel or $\Om_1$}	\\
 \frac{c}{3} \log \frac{|1-x_3^\alpha|\,|x_4^\alpha - x_5^\alpha|}{|1-x_5^\alpha|\,|x_3^\alpha - x_4^\alpha|}	&\text{for $t$-channel or $\Om_3$}	
\end{dcases*}
\end{align}
\section{Conformal blocks and entanglement entropy from holography}\label{sec:5}

In this section, we shall reproduce the conformal blocks considered in Section \ref{sec:3} from the gravity dual. This will involve a simple generalization of the bulk picture developed in \cite{Hijano:2015rla}. 
As we had noticed before,  the heavy operators in the CFT creates an excited state. In the bulk, this state can be thought in terms of a deformation of global AdS$_3$ into a conical defect geometry. 
From the conventional holographic dictionary, the  primaries in the CFT are dual scalar fields in the bulk. However, since the conformal dimension of these operators scale as the central charge, the mass of the bulk scalar also scales as $c$ ($M=\sqrt{h_L(h_L-1)} \sim c \gg 1$) and can be approximated by worldlines of point-particles. 
It was shown  in \cite{Hijano:2015rla}, that the momenta along these worldlines are equal to the accessory parameters of the conformal block.

We work with asymptotically AdS$_3$ space in the global co-ordinates in which the dual CFT lives on a cylinder.  
The metric of the geometry dual to the heavy state is given by \cite{Hijano:2015rla,Alkalaev:2015wia,Alkalaev:2015fbw}
\begin{eqnarray}\label{metric}
 ds^2 = \frac{\alpha^2}{\text{cos}^2 \rho} \left( -dt ^2 + \frac{1}{\alpha^2} d\rho^2 +\text{sin}^2 \rho \, d\phi^2 \right),  \quad \text{with }\alpha = \sqrt{1-24h_H/c}.
\end{eqnarray}
Depending on whether $\alpha^2 > 0$ or $\alpha^2 < 0$ the metric represents a conical defect with the singularity at $\rho =0$
or  a BTZ black hole with the event horizon at $\rho =0$, respectively. The boundary is at $\rho = \frac{\pi}{2}$. To avoid
potential divergences, we use the regularization $\text{cos} \,\rho|_{\rho\to\frac{\pi}{2}} = \Lambda^{-1}$ (where $\Lambda$ is the UV cutoff in the dual field theory for momenta or energies).  
We shall work with a constant time slice of \eqref{metric} which is a disc with radial and
angular variables as $\rho$ and $\phi$ respectively. 

The motion of the particle on the background above 
is described by the worldline action 
\begin{align}\label{worldline-action}
 S = M  \int_{\lambda_i}^{\lambda_f} d\lambda \sqrt{ g_{t t} \dot{t}^2+ g_{\rho \rho} \dot{\rho}^2 + g_{\phi \phi} \dot{\phi}^2}  \ . 
\end{align}
We restrict our attention to the constant time slice  $\dot{t} = 0$. 
The required geodesic segments between any two points on the disc can be obtained by extremizing the 
worldline action \eqref{worldline-action}. If we choose the parameter $\lambda$ as the proper length, the geodesic equation reads  \cite{Alkalaev:2015wia,Hijano:2015rla}
\begin{align}
 \frac{1}{\text{cos}^2 \rho}  \dot{\rho}^2 + \frac{p_{\phi}^2}{\alpha^2} \text{cot}^2 \rho = 1\ ,
\end{align} 
where, $p_{\phi}= \a^2 \ \text{tan}^2 \, \rho \, \dot{\phi}$, is the conserved momentum conjugate to $\phi$.
The solution to the geodesic equation is 
\begin{align}
\text{cos}\, \rho = \frac{1}{\sqrt{1 + p_{\phi}^2/\a^2}} \frac{1}{\text{cosh} \, \lambda}  \ ,
\end{align}
 Using these relations one can compute the regularized length of geodesics which will be shown to reproduce the corresponding
 conformal blocks.

In what follows, we illustrate the worldline 
configurations corresponding to the 3-point function and the 4-point conformal block in the CFT. Then we shall generalize
the bulk picture for conformal blocks with arbitrary number of odd and even operator insertions.  
\subsection*{Worldlines corresponding to 3-point function}
Let us consider the bulk realization of the 3-point function $\langle \oh(\infty) \ol(z) \oh(0) \rangle $. This is effectively a 1-point function in an excited state.  In terms of the cylinder coordinates on the CFT, $(w =-i\, \log\, z)$, this is realized in the bulk as a radial geodesic from the point of insertion of the  {light}
operator to the singularity (see \fig{3-4}). The corresponding conformal
block can be computed by determining the regulated length of the  geodesic from the position of defect  $\rho = 0$ to the
boundary i.e, $\text{cos} \, \rho|_{\rho \to \pi/2} = \Lambda^{-1}$.
\begin{align}
 l_L = \int_{0}^{\text{cos} \, \rho = \Lambda^{-1}} \frac{d \rho}{\text{cos} \, \rho} = \log\left[\frac{\sin \left(\frac{\rho }{2}\right)+\cos \left(\frac{\rho }{2}\right)}{\cos \left(\frac{\rho }{2}\right)-\sin \left(\frac{\rho }{2}\right)}\right]_{0}^{\text{cos} \, \rho = \Lambda^{-1}} = \log (2 \Lambda) + \mathcal{O}(\Lambda^{-2}) \ .
\end{align}
The contribution to the correlator is given by
\begin{align}
 G(w) = e^ {-h_L \, l_L} \approx (2 \Lambda)^{-h_L}\ .
\end{align}
This function is independent of $w$ and just depends on the cut-off $\Lambda$. However, this is still a result on the cylinder. To obtain the conformal 
block on the plane, we use the standard exponential map  $z = e^{i w}$ to obtain
\begin{align}\label{bulk3}
 \mathcal{G}(z) = z^{-h_L} G(w)\big |_{w=-i\, \log\, z} = (2 \Lambda z)^{-h_L} \propto z^{-h_L} \ .
\end{align}
There are no additional bulk worldlines possible in this case unlike the higher point functions as we shall see below. 
Equation \eqref{bulk3} precisely reproduces the $z$ dependence of the normalized 3-point function, $ \langle \oh(\infty) \ol(z) \oh(0) \rangle/ \langle \oh(\infty) \oh(0) \rangle  $, which is fixed by conformal invariance. 
\begin{figure}[!t]
	\centering
	\begin{tabular}{c}
		\includegraphics[width=.7\textwidth]{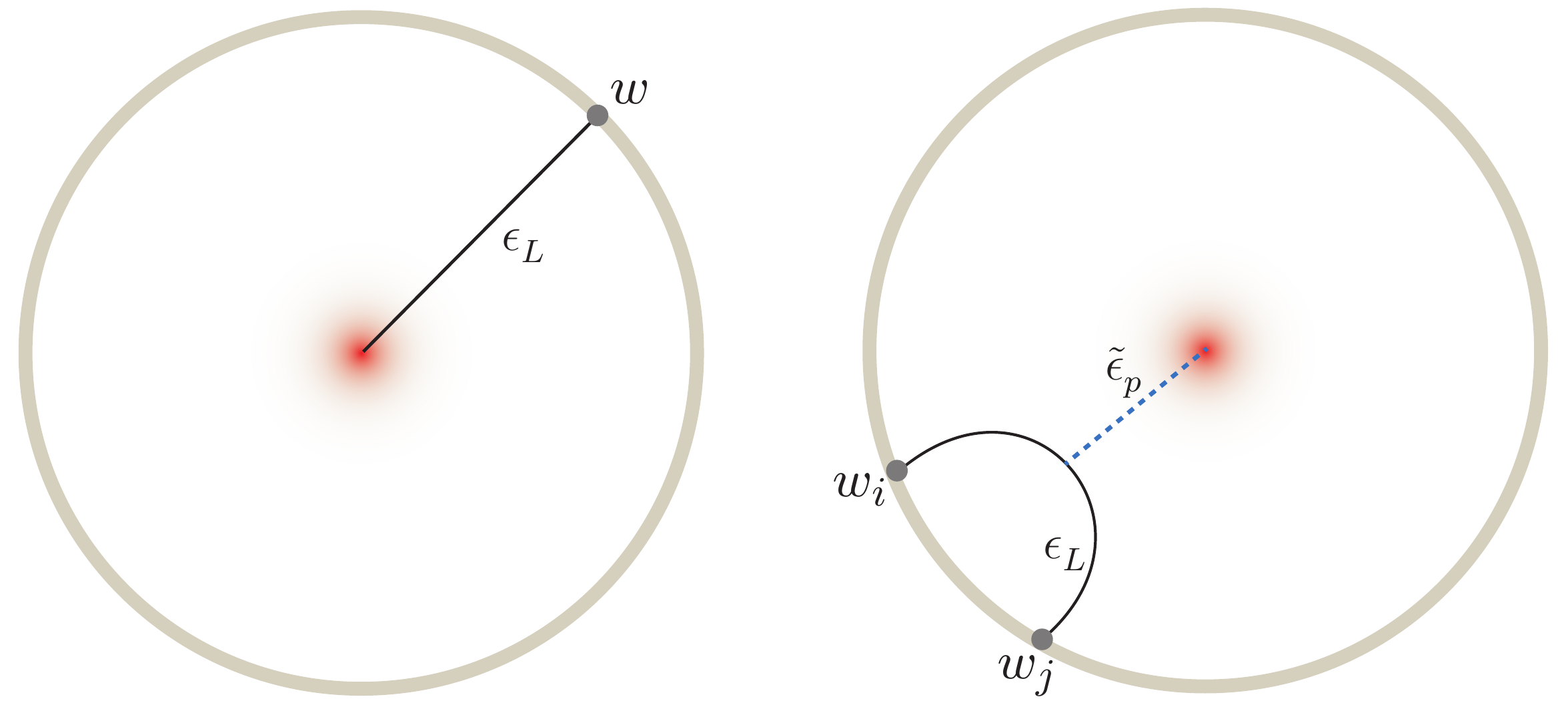} 
		\end{tabular}
		\caption{\small Bulk picture of the 3-point function (\textit{left})  and 4-point conformal block (\textit{right}) in CFT. Two of the operators are  {heavy} which deform the 
		background geometry (from the vacuum AdS to the conical defect)  and the other {light} operators are described by geodesics of massive probe particles.}
		\label{3-4}
		\end{figure}
\subsection*{Worldlines corresponding to 4-point conformal block}

The conformal block of the 4-point function $\langle \oh(\infty) \ol(x_i) \ol(x_j) \oh(0) \rangle $ is described in the bulk by the geodesic configuration shown in Fig.\,\ref{3-4} \cite{Hijano:2015rla}. 
The configuration consists of a geodesic anchored at 
the points of insertion of light operators, namely $w_i$ and $w_j$ on the cylinder\footnote{This geodesic is same as the Ryu-Takayanagi one.}. In addition to this, there is another geodesic -- which represents exchange of primary operator with conformal dimension $\tilde{h}_p$ -- 
stretched between the singularity ($\rho =0$) and the former geodesic (see \fig{3-4}). The point of intersection of the geodesic segments
can be determined by minimizing the worldline action \cite{Hijano:2015rla}. For the case we are considering, in which the two light operators have same conformal  dimensions, the
 dotted worldline joins the mid-point of the geodesic connecting $w_i$ and $w_j$ (see \fig{3-4}). Therefore, the worldline
action becomes 
\begin{align}
 S =  \epsilon_L l_L + \tep _p l_p , 
\end{align}
Here, $l_L$ is the length of the geodesic joining the light operators at the boundary whilst $l_p$ is the length of the other geodesic joining the singularity and the mid-point of the former geodesic.
We also assume $\tep_p \ll \epsilon_L$, such that the radial geodesic does not backreact to the other one \cite{Hijano:2015rla}. Recall, that in our CFT analysis, the same assumption $\tep_p \ll \epsilon_L$ led to the dominant contribution by heavy operators in the horizontal intermediate channels.
 For the 
$\epsilon_L$ geodesic (here, $w_{ij}=w_i-w_j$), we have
\begin{align}
 \text{cos} \, \rho = \frac{\text{sin} \,(\frac{\a w_{ij}}{2})}{\text{cosh} \, \lambda}
\end{align}
The regulated $\epsilon_L$ and $\epsilon_p$ geodesic lengths are given by
\begin{align}\label{RT-length}
  l_L (w_{ij}) &= 2 \lambda \big |_{\text{cos} \, \rho = \Lambda^{-1}} = 2 \, \log \left(\text{sin}\frac{\a w_{ij}}{2}\right) + 2 \, \log \left(\frac{\Lambda}{2}\right) ,
\\
\label{int-length}
  l_p (w_{ij}) &= \int_{0}^{\text{cos} \, \rho = \text{sin}\frac{\a w_{ij}}{2} } \frac{d \rho}{\text{cos} \, \rho} = - \, \log \left( \text{tan} \frac{\a w_{ij}}{4}  \right) .
\end{align}
The limits of integration for $l_p$ are the ones corresponding to the singularity and the mid-point of the $\ep_L $ geodesic. The ($w_i$ dependent) contribution to the correlator is given by \cite{Hijano:2015rla}
\begin{align}
 G(w_i ,w_j) &= e^ {-\frac{c}{6} S(w_i ,w_j)}  = e^ {- h_L \, l_L (w_{ij}) - \th_p  \, l_p(w_{ij})}  = \frac{\left( \text{tan} \frac{\a w_{ij}}{4}  \right)^{\th_p}}{\left(\text{sin}\frac{\a w_{ij}}{2} \right)^{2h_L} } .
\end{align}
Once again, to obtain the conformal block on the plane we need to perform the conformal transformations, 
$x_i = e^{i w_i}$ and $x_j = e^{i w_j}$
\begin{align}
 \mathcal{F}_\fear(x_i,x_j) = x_i^{-h_L} x_j^{-h_L}G(w_i,w_j)\big |_{w_{i,j}=-i\, \log\, x_{i,j}} .
\end{align}
The extra prefactor comes due to the conformal transformation of the light operator $\ol(x)$. 
%
 In terms of the  following function 
\begin{align}\label{eff}
f(x_i,x_j) 
 &= \left[ \epsilon_L \log  ( x_i x_j) + 2  \epsilon_L  \log \left(\text{sin} \frac{\a w_{ij}}{2}\right)- \tep_p   \log \left(\text{tan} \frac{\a w_{ij}}{4}\right)    \right]_{w_{i,j}=-i \log x_{i,j}} \nonumber \\
 &=   2 \, \epsilon_L \, \log \frac{ x_i^{\alpha}- x_j^{\alpha}  }{  (x_i x_j)^{\frac{\alpha-1}{2}} }  + \tep_p \, \log \frac{x_i^{\alpha /2}+ x_j^{\alpha /2}}{x_i^{\alpha /2}-x_j^{\alpha /2}} \ ,
\end{align}
the conformal block $ \mathcal{F}_{\fear}(x_i,x_j)$ is then given as
\begin{align}
\mathcal{F}_\fear(x_i,x_j)=\exp(-cf(x_i,x_j)/6) \ .
\end{align}
which indeed agrees with the CFT result \eqref{four-point-block} up to the constant $(\tep_p\log 4\alpha - 2 \ep_L \log \alpha)$.

\subsection*{\textit{Worldlines corresponding to higher-point conformal blocks}}
		
Largely motivated by the Ryu-Takayanagi proposal for disjoint intervals, the worldline configurations above can be extended to conformal blocks of arbitrary $(m+2)$-point functions of  two heavy operators and an arbitrary number $(m)$ of light operators
$$ \Big\langle \oh(\infty) \left[ \ol(1) \prod_{i=3}^{m +1}\ol (x_i) \right] \oh(0) \Big\rangle  $$ 
The basic constituents of the holographic representation of these higher conformal blocks are the worldline configurations for the 3-point function and the 4-point block. 
One then needs to add up lengths of the geodesic segments in the bulk ($l_v$), weighted with the corresponding scaled conformal dimensions ($\ep_v$), to obtain the  worldline action  
\begin{eqnarray}
S(w_i)= \sum_{v} \ep_v l_v
\end{eqnarray}
and the conformal block is given by $\exp[-\tfrac{c}{6}S(w_i)]$.
Additionally, we need to perform a  conformal transformation to go from the cylinder to the plane. Like the 3- and 4-point examples considered above, we shall mostly concern ourselves with the dependence of conformal block on the cross-ratios ($x_i$).
We now describe the bulk pictures of the two distinct cases.  
 \begin{figure}[!t]
 	\centering
 	\begin{tabular}{c}
 		\includegraphics[width=.9\textwidth]{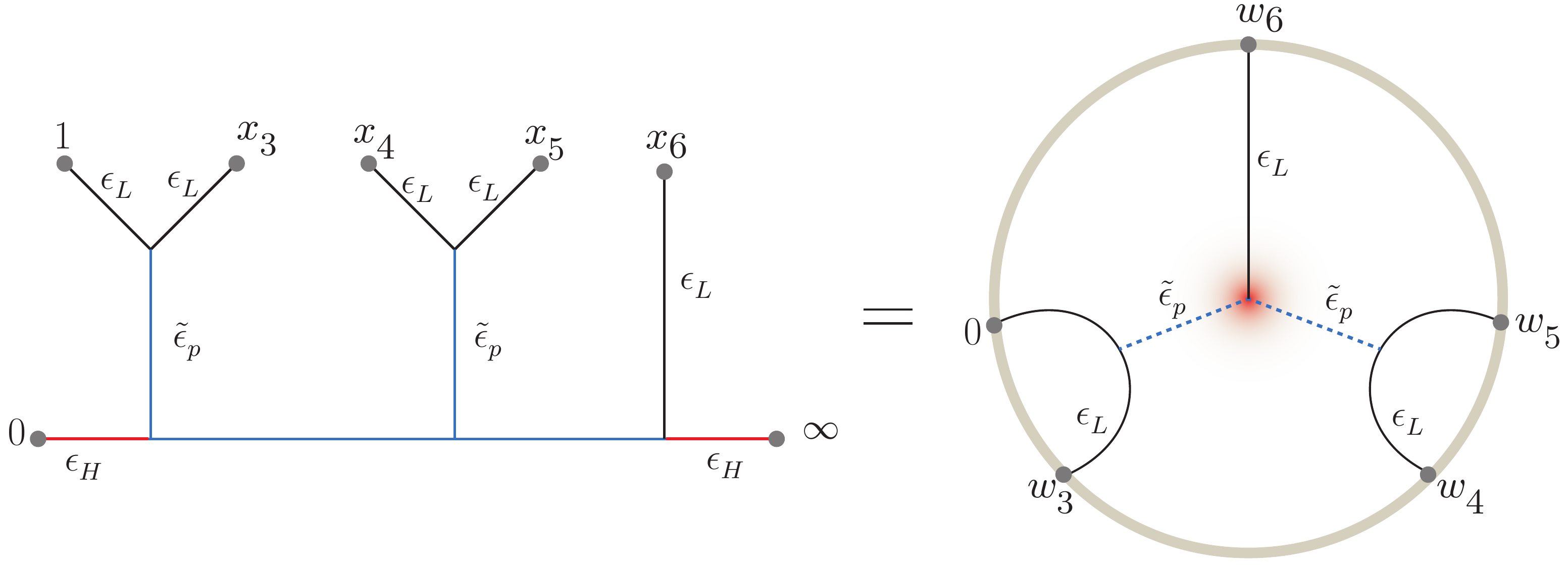} 
 	\end{tabular}
 	\caption{\small Worldline configuration corresponding to $[(2k+1)+2]$-point conformal block (with $k=2$ above). There are $k$ geodesics connecting
 		$2k$ points of light insertions of the CFT  whereas one geodesic segment connects the remaining point of insertion to the singularity.
 		Also, there are intermediate exchanges described by the dotted geodesics from the singularity to mid points of the boundary-to-boundary
 		geodesics.}
 	\label{odd}
 \end{figure}
\subsubsection*{Odd number of light insertions}		

Let us consider a $[(2k+1)+2]$-point function in the CFT where two of them are heavy operators whereas an odd number $2k+1$ of them are light ($k \in \mathbb{N}$). Generalizing the
previous bulk descriptions, we can see that amongst these $2k+1$ points, $2k$  points of light operator insertions will pairwise form $k$
geodesic segments (see \fig{odd}). 
These worldlines joining a pair of points in the boundary will also have an additional geodesic segment representing the intermediate exchange of primaries of dimension $\tilde{h}_p$ whose common origin is the singularity
and each one ends on the mid-points of the boundary-to-boundary geodesic segments.
There is another geodesic segment originating from the remaining (or unpaired) light operator insertion will anchor into the singularity at the centre ($\rho =0$). 
This is precisely the factorization of $[(2k+1)+2]$ point conformal block 
 into $k$ 4-point blocks and a 3-point function. 
 After summing the geodesic lengths, the net contribution to correlator is (here, $|j|=\prod_{s=3}^{m+1}x_s^{-h_L}$ is the factor arising from the conformal transformation from the cylinder to  the plane $x_s=e^{iw_s}$)
 \begin{align}\label{bulk-factorization-odd}
 \cF_{((2k+1)+2)}=\ &|j|G(w_i(x_i)) = |j| e^ {-\frac{c}{6} S(w_i(x_i))} = |j| e^ {- h_{L}  \, l_{L_a}}  \prod_{\Om_i \mapsto \lb (i,j) \rb}  e^ {-  h_{L } \, l_{L}(w_{ij})}  e^{- \tilde{h}_{p}  \, l_{p}(w_{ij})} \nn \\ =\ &x_a^{-h_L}\prod_{\Om_i \mapsto \lb (i,j) \rb} \cF_{\fear}(x_i,x_j) \ .
\end{align} 
The prefactor $x_a^{-h_L}$ represents any one of the points of insertion which is left over after  connecting the  others pairwise by the boundary-to-boundary geodesic
 segments. The choice of pairings via the geodesics in the bulk is in one-to-one correspondence with OPE channels,  $\Om_i$, in the CFT\footnote{It may be of potential concern that the worldlines in channels other than the $s$-channel may intersect each other. Such issues can be avoided by suitably considering infinitesimally separated  constant time slices each containing contributions for  4-point block(s) or a 3-point function.
 	In any case, the sum of lengths of geodesics will not change. }. 
 Therefore, the above equation \eqref{bulk-factorization-odd} precisely matches with the CFT result for odd-point blocks -- \eqref{big-block-odd1} or \eqref{big-block-odd2}. 

\begin{figure}[!t]
	\centering
	\begin{tabular}{c}
		\includegraphics[width=.99\textwidth]{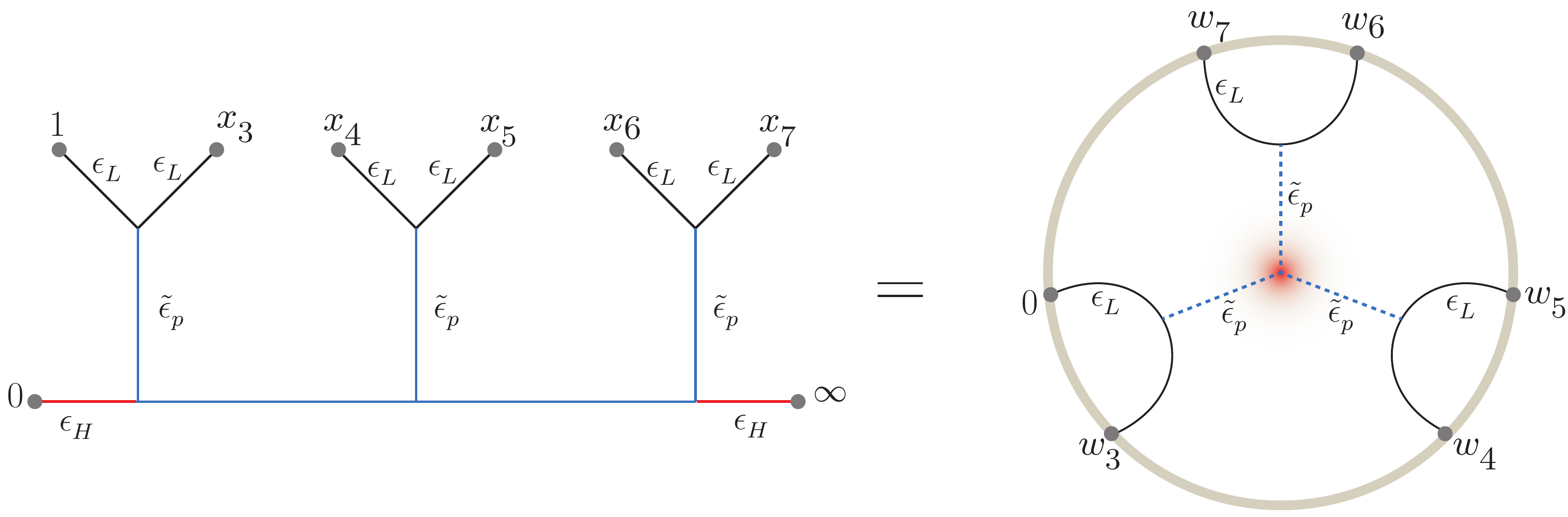} 
	\end{tabular}
	\caption{\small Worldline configuration corresponding to $(2k+2)$-point conformal block (with $k=3$). There are $k$ geodesics connecting
		$2k$ points of light insertions on the CFT. }
	\label{even}
\end{figure}
\subsubsection*{Even number of light insertions}
For an even number  ($2k$) of light operator insertions, we can form $k$ boundary-to-boundary geodesics joining a pair of light operator insertions ($k \in \mathbb{N}$).  
Again there 
are additional worldlines starting from the singularity and ending on the mid-point of each boundary-to-boundary  geodesic which geometrically describe the intermediate  exchanges ($\th_p$).  This case is, therefore, effectively equivalent to the odd-point block upon removal of the extra geodesic from the unpaired light insertion. The contribution
to the correlator is given once again in terms of the worldline action as  
\begin{align}
  \cF_{(2k+2)}= \ &|j|G(w_i(x_i)) = |j| e^ {-\frac{c}{6} S(w_i(x_i))} = |j|  \, \prod_{\Om_i \mapsto \lb (i,j) \rb}  e^ {-  h_{L } \, l_{L}(w_{ij})}  e^{- \tilde{h}_{p}  \, l_{p}(w_{ij})} \nn \\ =\ & \prod_{\Om_i \mapsto \lb (i,j) \rb} \cF_{\fear}(x_i,x_j) \ .
\end{align}
(The factor $|j|$ used here, once again, arises from conformal transformations from the cylinder to the plane and is given by $\prod_{s=3}^{m+1}x_s^{-h_L}$.)
This agrees with the CFT result for even point blocks obtained in \eqref{big-block}.

It can therefore be seen that, the conformal blocks in the OPE channels we have considered bears a very natural interpretation in terms of bulk worldline diagrams. Furthermore, the network of geodesics considered here are simpler compared to those in \cite{Alkalaev:2015fbw,Alkalaev:2015wia} which correspond to other OPE channels. Evidently, the basis, which we have chosen to work with, admits the straightforward generalization to an arbitrary number of light operator insertions.

\subsection*{\textit{Holographic entanglement entropy}}
The Ryu-Takayanagi prescription \cite{Ryu:2006bv}, prescribes that the entanglement entropy is given by the minimal area of a surface in AdS anchored at the endpoints of the interval(s) 
\begin{align*}
S_\cA = \frac{\min[\gamma_\cA]}{4G_N } \ .
\end{align*}
For the case of AdS$_3$, the minimal area surface is equivalent to a geodesic. For the case of multiple intervals, there are several of these geodesic configurations, $\mathbf{G}_i$, which are possible in the bulk, out of which we need to choose the one with the minimal length. Furthermore, $G_N$ is related to the central charge by the Brown-Henneaux relation $c=3\ell/2G_N$ \cite{brown1986}. 

 For the metric \eqref{metric}, the length of the geodesic joining two points in the boundary (CFT on the cylinder) has been calculated in \eqref{RT-length} -- see also \cite{Roberts:2012aq} and Appendix A of \cite{Hartman2}. Considering the CFT on the boundary to be living on a plane, the length of the bulk geodesic joining the points $x_i$ and $x_j$ (both real) is
\begin{align}
l_{ij}= 2 \, \log \frac{ x_i^{\alpha}- x_j^{\alpha}  }{\alpha  (x_i x_j)^{\frac{\alpha-1}{2}} }  \ .
\end{align}
Hence, summing over all the geodesics and applying the minimal-area condition, the result for holographic entanglement entropy is 
\begin{align}
S_\cA = \frac{c}{3}\min_i \Bigg\lb \sum_{ \mathbf{G}_i \mapsto  \lb (p,q) \rb} \log \frac{ x_p^{\alpha}- x_q^{\alpha}  }{ \alpha (x_p x_q)^{\frac{\alpha-1}{2}} }  \Bigg\rb \ .
\end{align}
This agrees exactly with \eqref{ee-disjoint} -- provided   the bulk geodesic configurations $\mathbf{G}_i$ are identified with the OPE channels $\tom_i$ in the CFT (cf.~\cite{Headrick:2010zt,Hartman1,Faulkner} for the vacuum case). Hence, depending on the values of the cross-ratios $x_i$ the relevant OPE channel is chosen in the CFT and  analogously the geodesic configuration of minimal length is the one that reproduces the corresponding entanglement entropy of the heavy excited state\footnote{See \cite{Caputa:2014vaa,Nozaki:2014hna,He:2014mwa,Nozaki:2014uaa,Guo:2015uwa,Caputa:2015tua,Chen:2015usa,Mukherjee:2014gia} for other results on  entanglement entropy  in excited states. }.

\section{Moduli space of the correlation function}
\label{sec:7}
As remarked earlier, the correlation function of $m$ light operators and 2 heavy operators on the plane is associated with the Riemann sphere with $(m+2)$ punctures, $\Sigma_{0,m+2}$. This can also be seen by thickening the diagrams of the conformal blocks. The expansion in terms of conformal partial waves is, then, the decomposition of this Riemann surface into 3-holed spheres. The moduli space of $\Sigma_{0,m+2}$ is $\cM_{0,m+2}$ which has $(m-1)$ complex moduli $(x_i,\bar{x}_i)$ \cite{Moore:1988qv,Zwiebach:1990nh}.  

The OPE channels, which involve the pairwise fusion of two light operators, located at $1,x_{3},\cdots,x_{m+1}$, describe the moduli space around some specific regions. For instance, the 6-point function in the $s$-channel, $\Om_1$, is restricted to the disjoint regimes around $x_3 \to 1$ and $x_4 \to x_5$. Whereas, the $t$-channel, $\Om_2$, describes the region around $x_5 \to 1$ and $x_3 \to x_4$. The $u$-channel or  $\Om_3$ describes the region around $x_4 \to 1$ and $x_3 \to x_5$. The three worldline configurations  are the bulk duals  equivalently describing these  OPE channels. In general, at large central charge, the $(m+2)$-point correlation function   switches from regions  of one OPE channel of  $\cM _{0,m+2}$ to another in the moduli space upon tuning the cross-ratios\footnote{It is worthwhile to note, that this jump is seemingly discrete since we are strictly working in the $c\to\infty$ limit. In fact, it can be explicitly shown that quantum corrections to mutual information smoothen this discrete jump \cite{Barrella:2013wja}. We are grateful to Arnab Rudra for pointing this possibility.  }. Furthermore, the  worldline configurations in the bulk also correspond to each of the possible ways of decomposing the punctured Riemann sphere into 3-holed spheres. 

In the context of entanglement entropy of $N$ disjoint intervals, the correlator has $N$ twist and $N$ anti-twist operators in addition to two heavy operators. It was shown in \cite{Faulkner}, that the number of Ryu-Takayanagi geodesic configurations, $\cN_N$, is given by the recursive formula
\begin{align}
\cN_{N} = 3\cN_{N-1} - \cN_{N-2} \ .
\end{align}
Interestingly, $\cN_N$ above is given by the alternating Fibonacci numbers, $F_{2N-1}$ \cite{oeis}. As we had seen, the geodesic configurations $\mathbf{G}_i$ are in one-to-one correspondence with the contours considered in the monodromy problem, $\tom_i$. In fact, the contours in the CFT are smoothly shrinkable into the bulk without ever crossing its corresponding geodesic \cite{Faulkner}. Therefore, $F_{2N-1}$ also counts the number of possible OPE channels of the conformal blocks (equaling the number of allowed pant-decompositions of the $(2N+2)$-punctured Riemann sphere) in the basis of pairwise fusion of the twist and anti-twist operators. 
Each of these OPE channels has its regime of validity in $N$ disconnected regions of the moduli space, $x_{2i+1} \to 1$ and $x_{2j} \to x_{2k+1}$\footnote{The twist operators (located at 1 and $x_{\text{even}}$)  have non-vanishing OPEs with anti-twist operators (located at $x_{\text{odd}}$).  The OPE of (anti-)twist with itself is zero. Moreover, the location of the endpoints of the intervals are $1<x_3<x_4<\cdots<x_{m-1}$. This is the reason why the number of OPE channels gets reduced to $F_{2N-1}$ from $\nu^{\text{(even)}}_{2N} = (2N)!/(2^{N}N!)$.}.

\section{Conclusions}
\label{sec:8}
In this work, we have studied higher-point conformal blocks of two heavy operators and an arbitrary number of light operators in a CFT with large central charge. We focused our attention to a specific class of OPE channels in which  the light operators fuse in pairs and the conformal dimension of the operator in the intermediate channels (after fusion of two $\ol$s) are the same. In the heavy-light limit, we have been able to show that these blocks factorize into products of 4-point blocks. This was achieved using the monodromy method and the exponentiation of the block at large central charge. These CFT results could be reproduced from bulk worldline configurations using the methods presented in \cite{Hijano:2015rla}. We have also applied these results to study the entanglement entropy of an arbitrary number of disjoint intervals for excited states. Hence, this work serves as a twofold generalization of the results of \cite{Hartman1,Headrick:2010zt} to excited states and that of \cite{Hartman2} to multiple intervals. 
If further information about the spectrum (like structure constants and operator content)  of large-$c$ theories is available, these results can be possibly used to know higher-point heavy-light correlation functions in holographic CFTs\footnote{See \cite{Hartman:2014oaa,Keller:2014xba} for some progress along these lines.}. 

It would be interesting to find subleading corrections both in the light parameter $\ep_L$ as well as in  $1/c$ to the higher point conformal block \cite{Fitzpatrick:2015dlt,Beccaria:2015shq}. The $1/c$ corrections would lead to corrections in entanglement entropy as well. One can then try to see whether these can be holographically reproduced by considering one-loop determinants in handlebody geometries obtained from orbifolding the conical defect \cite{Barrella:2013wja}. Moreover, it can then be seen how $1/c$ effects smoothen the jumps in mutual information. 

An immediate application of our results on higher-point conformal blocks is to study time-dependent entanglement entropy of disjoint intervals and mutual information in local quenches. The evolution of entanglement entropy  of a single interval in this scenario has been studied in \cite{Hartman2} and also the mutual information for joining quenches have been studied in \cite{Asplund:2013zba}. 

As we had mentioned in the introduction, there is a fascinating connection between conformal blocks (of a specific pant decomposition) and Nekrasov partition functions arising from the AGT correspondence \cite{Alday:2009aq,Wyllard:2009hg,Nekrasov:2002qd}. This relation was utilized in \cite{Alkalaev:2015wia} to derive results for heavy-light blocks. Although, their conformal blocks are in a different basis and explicit results exist only for the 4- and 5-point blocks, it would be interesting to precisely relate the results of \cite{Alkalaev:2015wia} to those found here via fusion transformations. Furthermore, it also known that the one-point function of chiral ring elements in the $4d$ gauge theory are related to CFT$_2$ conformal blocks with insertions of conserved charges  \cite{Alba:2010qc}. This  presents the exciting prospect of utilizing this connection to find the entanglement entropy of heavy states in presence of chemical potentials\footnote{We are grateful to Sujay Ashok for making us aware of this and for related discussions.}. 

Another intriguing direction  is to see how the above results for higher-point conformal blocks generalize to theories with additional conserved currents -- these include supersymmetric and higher-spin extensions. Our analysis   suggests that the factorization could happen for conformal blocks of these theories as well. For the case of CFTs which have higher-spin gravity duals, the bulk Wilson line prescription \cite{Hijano:2015qja,deBoer:2014sna} may  also suggest such a factorization of the higher-point blocks. 


Finally, it would be exciting to see to what extent our analysis of conformal blocks have analogues in higher dimensions. From the holographic side, one can make use of the geodesic Witten diagram prescription of \cite{Hijano:2015zsa}. In a related context, correlation functions of heavy and light operators have also been considered in the context of $\cN=4$ super Yang-Mills and also from the dual string theory \cite{Costa:2010rz,Zarembo:2010rr,Bajnok:2014sza,Escobedo:2011xw,Hollo:2015cda}. The heavy operators  correspond to classical string solutions and the typically protected light operators correspond to supergravity modes. The heavy states can be expressed in terms of Bethe states in the spin-chain description of the planar limit. The correlation function is then reduced to finding expectation values of light operators in these states. Although the existing  results (involving a sufficiently intricate analysis) are mostly for the structure constant $c_{\text{HHL}}$ appearing in the 3-point function, it would be interesting to see a whether a higher point generalization of the same shares any features with its lower dimensional counterpart considered here. 

\section*{Acknowledgements}

S.D would like to thank Justin David and Juan Jottar for fruitful discussions on several aspects of this problem. We are grateful to Sujay Ashok for discussions on conformal blocks from the  AGT correspondence  and valuable comments on our manuscript.  We thank Roji Pius and Arnab Rudra for clarifying aspects of Riemann surfaces and their moduli spaces. We also thank  Chi-Ming Chang, Diptarka Das, Yunfeng Jiang, Maximilian Kelm, Christoph Keller, Sandipan Kundu, B.~Sathiapalan, Ying Lin,  Amit Sever and Xi Yin for crucial discussions. 
We thank Pawel Caputa, Matthias Gaberdiel and Gautam Mandal for discussions and suggestions on the manuscript.
R.S would like to thank the `Indo-Israel Strings Conference 2015, Goa, India', where the work was first presented. S.D thanks the organizers of `Iberian Strings 2016' at IFT Madrid for an opportunity to present this work.  The work of S.D is supported by the NCCR SwissMap, funded by the Swiss National Science Foundation.

\providecommand{\href}[2]{#2}

%
%
%
%
%

\end{document}